\documentstyle[12pt]{article}

\catcode`\@=11
\def\marginnote#1{}
\newcount\hour
\newcount\minute
\newtoks\amorpm
\hour=\time\divide\hour by60 \minute=\time{\multiply\hour by60
\global\advance\minute by-\hour}\edef\standardtime{{\ifnum\hour<12
\global\amorpm={am}%
        \else\global\amorpm={pm}\advance\hour by-12 \fi
        \ifnum\hour=0 \hour=12 \fi
        \number\hour:\ifnum\minute<10
0\fi\number\minute\the\amorpm}}
\edef\militarytime{\number\hour:\ifnum\minute<10
0\fi\number\minute}

\def\draftlabel#1{{\@bsphack\if@filesw {\let\thepage\relax
   \xdef\@gtempa{\write\@auxout{\string
      \newlabel{#1}{{\@currentlabel}{\thepage}}}}}\@gtempa
   \if@nobreak \ifvmode\nobreak\fi\fi\fi\@esphack}
        \gdef\@eqnlabel{#1}}
\def\@eqnlabel{}
\def\@vacuum{}
\def\draftmarginnote#1{\marginpar{\raggedright\scriptsize\tt#1}}
\def\draft{\oddsidemargin -.5truein
        \def\@oddfoot{\sl preliminary draft \hfil
        \rm\thepage\hfil\sl\today\quad\militarytime}
        \let\@evenfoot\@oddfoot \overfullrule 3pt
        \let\label=\draftlabel
        \let\marginnote=\draftmarginnote

\def\@eqnnum{(\theequation)\rlap{\kern\marginparsep\tt\@eqnlabel}%
\global\let\@eqnlabel\@vacuum}  }

\def\numberbysection{\@addtoreset{equation}{section}
        \def\theequation{\thesection.\arabic{equation}}}

\def\underline#1{\relax\ifmmode\@@underline#1\else
 $\@@underline{\hbox{#1}}$\relax\fi}

\catcode`@=12 \relax

\numberbysection

\topmargin by -\headsep

\def\br{\begin{eqnarray}}
\def\er{\end{eqnarray}}
\def\be{\begin{equation}}
\def\ee{\end{equation}}

\def\lb{\lbrack}
\def\rb{\rbrack}

\def\({\left(}
\def\){\right)}

\relax

\newcommand\sbr[2]{\left\lbrack\,{#1}\, ,\,{#2}\,\right\rbrack}

\def\a{\alpha}

\def\d{\delta}

\def\g{\gamma}

\def\h{{1\over 2}}
\def\l{\lambda}
\def\L{\Lambda}

\def\pa{\partial}

\def\s{\sigma}

\def\tp0{\Theta_{+}^{(0)}}
\def\tm0{\Theta_{-}^{(0)}}

\def\vp{\varphi}

\def\f#1#2#3 {f^{#1#2}_{#3}}

\def\win1{{\sf w_{1+\infty}}}

\def\Win1{{\sf W_{1+\infty}}}

\def\rlx{\relax\leavevmode}
\def\inbar{\vrule height1.5ex width.4pt depth0pt}
\def\IZ{\rlx\hbox{\sf Z\kern-.4em Z}}
\def\IR{\rlx\hbox{\rm I\kern-.18em R}}
\def\IC{\rlx\hbox{\,$\inbar\kern-.3em{\rm C}$}}
\def\IN{\rlx\hbox{\rm I\kern-.18em N}}
\def\IO{\rlx\hbox{\,$\inbar\kern-.3em{\rm O}$}}
\def\IP{\rlx\hbox{\rm I\kern-.18em P}}
\def\IQ{\rlx\hbox{\,$\inbar\kern-.3em{\rm Q}$}}
\def\IF{\rlx\hbox{\rm I\kern-.18em F}}
\def\IG{\rlx\hbox{\,$\inbar\kern-.3em{\rm G}$}}
\def\IH{\rlx\hbox{\rm I\kern-.18em H}}
\def\II{\rlx\hbox{\rm I\kern-.18em I}}
\def\IK{\rlx\hbox{\rm I\kern-.18em K}}
\def\IL{\rlx\hbox{\rm I\kern-.18em L}}
\def\one{\hbox{{1}\kern-.25em\hbox{l}}}
\def\0#1{\relax\ifmmode\mathaccent"7017{#1}%
B        \else\accent23#1\relax\fi}

\def\EPJC#1#2#3{{\sl Eur. Phys. J.} {\bf C#1} (#2) #3}
                \def\JHEP#1#2#3{{\sl JHEP} {\bf#1} (#2) #3}
                \def\PRL#1#2#3{{\sl Phys. Rev. Lett.} {\bf#1} (#2) #3}
                \def\NPB#1#2#3{{\sl Nucl. Phys.} {\bf B#1} (#2) #3}

                \def\PRD#1#2#3{{\sl Phys. Rev.} {\bf D#1} (#2) #3}
                
                \def\PLA#1#2#3{{\sl Phys. Lett.} {\bf #1A} (#2) #3}
                \def\PLB#1#2#3{{\sl Phys. Lett.} {\bf #1B} (#2) #3}
                \def\JMP#1#2#3{{\sl J. Math. Phys.} {\bf #1} (#2) #3}

                \def\AoP#1#2#3{{\sl Annals Phys.} {\bf #1} (#2) #3}
                
                \def\RMP#1#2#3{{\sl Rev. Mod. Phys.} {\bf #1} (#2) #3}
                \def\PR#1#2#3{{\sl Phys. Reports} {\bf #1} (#2) #3}

                \def\LMP#1#2#3{{\sl Letters in Math. Phys.} {\bf #1} (#2) #3}

                \def\TMP#1#2#3{{\sl Theor. Mat. Phys.} {\bf #1} (#2) #3}
                \def\JPA#1#2#3{{\sl J. Physics} {\bf A#1} (#2) #3}

                \def\JPIV#1#2#3{{\sl J. Phys. IV} {\bf #1} (#2) #3}

                \def\a{\alpha}

                \def\d{\delta}

                \def\g{\gamma}
                
                \def\vp{\varphi}
                
                \def\h{ {1\over 2}  }

                \def\/{\frac}

                \def\l{\lambda}
                \def\L{\Lambda}

                \def\pa{\partial}

                \def\vp{\varphi}
                
                \def\s{\sigma}

                \def\({\Big(}
                \def\){\Big)}
                \def\[{\Big[}
                \def\]{\Big]}

                \def\rlx{\relax\leavevmode}
                \def\inbar{\vrule height1.5ex width.4pt depth0pt}
                \def\IZ{\rlx\hbox{\sf Z\kern-.4em Z}}
                \def\IR{\rlx\hbox{\rm I\kern-.18em R}}
                \def\IC{\rlx\hbox{\,$\inbar\kern-.3em{\rm C}$}}
                \def\IN{\rlx\hbox{\rm I\kern-.18em N}}
                \def\IO{\rlx\hbox{\,$\inbar\kern-.3em{\rm O}$}}
                \def\IP{\rlx\hbox{\rm I\kern-.18em P}}
                \def\IQ{\rlx\hbox{\,$\inbar\kern-.3em{\rm Q}$}}
                \def\IF{\rlx\hbox{\rm I\kern-.18em F}}
                \def\IG{\rlx\hbox{\,$\inbar\kern-.3em{\rm G}$}}
                \def\IH{\rlx\hbox{\rm I\kern-.18em H}}
                \def\II{\rlx\hbox{\rm I\kern-.18em I}}
                \def\IK{\rlx\hbox{\rm I\kern-.18em K}}
                \def\IL{\rlx\hbox{\rm I\kern-.18em L}}
                \def\one{\hbox{{1}\kern-.25em\hbox{l}}}
                \def\0#1{\relax\ifmmode\mathaccent"7017{#1}%
                B        \else\accent23#1\relax\fi}
                
\begingroup\makeatletter\ifx\SetFigFont\undefined
\def\x#1#2#3#4#5#6#7\relax{\def\x{#1#2#3#4#5#6}}%
\expandafter\x\fmtname xxxxxx\relax \def\y{splain}%
\ifx\x\y   
\gdef\SetFigFont#1#2#3{%
  \ifnum #1<17\tiny\else \ifnum #1<20\small\else
  \ifnum #1<24\normalsize\else \ifnum #1<29\large\else
  \ifnum #1<34\Large\else \ifnum #1<41\LARGE\else
     \huge\fi\fi\fi\fi\fi\fi
  \csname #3\endcsname}%
\else \gdef\SetFigFont#1#2#3{\begingroup
  \count@#1\relax \ifnum 25<\count@\count@25\fi
  \def\x{\endgroup\@setsize\SetFigFont{#2pt}}%
  \expandafter\x
    \csname \romannumeral\the\count@ pt\expandafter\endcsname
    \csname @\romannumeral\the\count@ pt\endcsname
  \csname #3\endcsname}%
\fi\endgroup
\begin{document}
\begin{titlepage}

                \begin{center}

                  {\large\bf Non-commutative solitons and strong-weak duality}

                \end{center}

\vspace{.5 cm}

                \begin{center}

                H. Blas$^{1}$, H. L. Carrion$^{2, 3}$ and M. Rojas $^{3}$\\

                \vspace{.6 cm}

               ${}^{1}$
               Departamento de Matem\'atica - ICET\\
Universidade Federal de Mato Grosso\\
 Av. Fernando Correa, s/n, Coxip\'o \\
78060-900, Cuiab\'a - MT - Brazil\\
$2$
Instituto de F\'{\i}sica, Universidade Federal do Rio de Janeiro,\\
Caixa Postal 68528, 21941-972 Rio de Janeiro, Brazil.\\
 ${}^{3}$
  {\small Centro Brasileiro de Pesquisas F\'\i sicas}\\ {\small Rua Dr.
                Xavier Sigaud, 150}\\

                {\small CEP 22290-180, Rio de Janeiro-RJ, Brazil}\\
  \end{center}










                \begin{abstract}

                \vspace{.4 cm}

                Some properties of the non-commutative versions of the sine-Gordon
                model (NCSG) and the corresponding massive Thirring
                theories (NCMT) are studied. Our method relies on the NC
                extension of integrable models and
                the master Lagrangian approach to deal with dual theories. The master
                Lagrangians turn out to be the NC versions of the so-called
                affine Toda model coupled to matter fields (NCATM) associated to the group
                $GL(2)$, in which
                the Toda field belongs to certain representations of either $U(1)\mbox{x} U(1)$  or  $U(1)_{C}$ corresponding to the Lechtenfeld et al. (NCSG$_{1}$) or Grisaru-Penati (NCSG$_{2}$) proposals
                for the NC versions of the sine-Gordon model, respectively.
                 Besides, the relevant NCMT$_{1, 2}$ models are written for
                 two (four)
                types of Dirac fields corresponding to the Moyal product
                extension of one (two) copy(ies) of the
                ordinary massive Thirring model. The NCATM$_{1,2}$ models share the same
                one-soliton (real Toda field sector
                 of model $2$) exact solutions, which are found without expansion in the NC
                 parameter $\theta$ for the corresponding Toda and matter fields describing
the strong-weak phases, respectively.
                 The correspondence NCSG$_{1}$ $\leftrightarrow$
                 NCMT$_{1}$ is promising since it is expected
                 to hold on the quantum level.

\end{abstract}





                \vspace{2 cm}




                \end{titlepage}


\section{Introduction}

\label{intro}

Field theories in non-commutative (NC) space-times are receiving
considerable attention in recent years in connection to the
low-energy dynamics of D-branes in the presence of background
B-field (see, e.g. Refs. \cite{seiberg}). In particular, the NC
versions of integrable systems (in two dimensions) are being
considered (see e.g. \cite{hamanaka}). It is believed that these
models, defined on two-dimensional NC {\sl Euclidean} space, turn
out to be the NC versions of statistical models in the critical
points and in the off-critical integrable directions.

Some non-commutative versions of the sine-Gordon model (NCSG) have
been proposed in the literature
\cite{grisaru1}-\cite{lechtenfeld}. The relevant equations of
motion have the general property of reproducing the ordinary
sine-Gordon equation when the non-commutativity parameter is
removed. The Grisaru-Penati version \cite{grisaru1, grisaru2}
introduces a constraint which is non-trivial only in the
non-commutative case. The constraint is required by integrability
but it is satisfied by the one-soliton solutions. However, at the
quantum level this model gives rise to particle production as was
discovered by evaluating tree-level scattering amplitudes
\cite{grisaru2}. On the other hand, introducing an auxiliary
field, Lechtenfeld et al. \cite{lechtenfeld} proposed a novel NCSG
model which seems to possess a factorizable and causal S-matrix.

Recently, in ordinary commutative space the so-called $sl(2)$
affine Toda model coupled to matter (Dirac) fields (ATM) has been
shown to be a Master Lagrangian (ML) from which one can derive the
sine-Gordon and massive Thirring models, describing the
strong/weak phases of the model, respectively
\cite{nucl}-\cite{nucl1}. Besides, the ML approach was
successfully applied in the non-commutative case to uncover
related problems in $(2+1)$ dimensions regarding the duality
equivalence between the Maxwell-Chern-Simons theory (MCS) and the
Self-Dual (SD) model \cite{ghosh}.

In this paper we  extend some properties of the ordinary $sl(2)$
ATM model to the NC case. We show that replacing the products of
fields by the $\star-$products, on the level of its effective
action,  the ATM theory is still an integrable field theory. Since
the ordinary effective action gives rise to equations of motion
which can be derived from a zero-curvature equation, we may
alternatively construct the NC extension of the ATM model directly
starting from its zero-curvature formulation. In this way the ATM
model belongs to those class of integrable field theories in which
the direct replacement of the $\star-$product in the action turns
out the model still integrable \cite{cabrera}. However, in our
case the NC extension of the WZW term in the ATM  effective action
must be considered properly.

The study of these models become interesting since the $su(n)$ ATM
theories constitute excellent laboratories to test ideas about
confinement \cite{nucl1, tension}, the role of solitons in quantum
field theories \cite{nucl}, duality transformations interchanging
solitons and particles \cite{nucl, jmp}, as well as the  reduction
processes of the (two-loop) Wess-Zumino-Novikov-Witten (WZNW)
theory from which the ATM models are derivable \cite{matter,
jhep}. Moreover,  the ATM type systems may also describe some low
dimensional condensed matter phenomena, such as self-trapping of
electrons into solitons, see e.g. \cite{brazovskii}, tunnelling in
the integer quantum Hall effect
 \cite{barci}, and, in particular, polyacetylene molecule systems in connection with
 fermion number fractionization \cite{jackiw}. It has been shown
 that the $su(2)$ ATM model describes the low-energy spectrum of QCD$_{2}$ ({\sl one flavor}
 and $N$ colors in the fundamental and $N=2$ in the adjoint representations,
 respectively)\cite{tension}.

 The paper is organized as follows. In the next section we present
 the NC extensions of the ATM model relevant to our discussions. This
 procedure deals with the choice of the group for the Toda field $g$. We introduce two
 types of master Lagrangians (NCATM$_{1,2}$), the {\sl first} one defined for $g \in U(1)\mbox{x}
 U(1)$ with the same content of matter fields as the ordinary ATM;
 the {\sl second} one defined for two copies of the usual ATM
 such that $g,\, \bar{g} \in U(1)_{C}$. In section \ref{ncsg12} the NCSG$_{1,2}$  models are derived from
 the relevant master Lagrangians through reduction procedures resembling the one performed in
 the ordinary ATM $\rightarrow$ SG reduction. In section \ref{decou} we decouple on shell the
theories NCSG$_{1,2}$ and
 NCMT$_{1,2}$, respectively. This procedure is further justified in sections
 \ref{ncmt12} and \ref{ncsol} when we consider the NCMT$_{1,2}$ Lagrangians and the soliton
 mappings satisfying the relevant decoupling equations. In section \ref{ncmt12} we consider the
 (bosonic) NCMT$_{1,2}$ models, as well as their global symmetries, associated currents and
 integrability properties. In these developments the double-gauging  of a U(1) symmetry in the
 star-localized Noether procedure to get the currents deserve a careful treatment. In section
 \ref{ncsol} we present the soliton
 solutions and establish the strong-weak duality between the NCSG$_{1,2}$ and NCMT$_{1,2}$ (real soliton sector of models $2$)
  models. The section \ref{concl} presents the conclusions and possible future directions. The Appendix A
  provides the affine $sl(2)$ Lie algebra
 properties. Some results of the ordinary ATM model
 are summarized in Appendix B.

 \section{The NC affine Toda coupled to matter (NCATM)$_{1, 2}$}

 \label{ncatm}

 The commutative Toda field $g$ in (\ref{fi1}) belongs to the complexified $U(1)_{C}$ group
 since in general $\vp \in \IC$. Different NC extensions of the ATM model (\ref{atmmat})
 are possible as long as all of them reproduce in the commutative limit the equations of
motion (\ref{atm1})-(\ref{atm3}).
 The symmetry group of the ordinary $SL(2)$ ATM model (see
 Appendices \ref{appa}  and \ref{appb}) in the NC case is not closed under $\star$; then, the
 NC extension requires the $GL(2)$ group. In the next steps we define two versions of
 the non-commutative $GL(2)$ affine Toda model
 coupled to matter (NCATM$_{1,2})$. Let us define the {\sl first} NC extension (NCATM$_{1}$) as
 \begin{eqnarray}
\nonumber
 S_{NCATM_{1}}& \equiv & S[g, W^{\pm},F^{\pm}]  \\\nonumber
 &=& I_{WZW}[g] + \int d^2x \{\frac{1}{2}<
\partial_{-} W^{-}\star [E_{2}\,,\, W^{-} ]> - \frac{1}{2}
< [E_{-2}\,,\,W^{+}] \star \partial_{+} W^{+} > +\\&& < F^{-}
\star \partial_{+} W^{+} > +<\partial_{-} W^{-}
                 \star F^{+}> + <F^{-}\star g \star F^{+} \star g^{-1}
                 >\},
               \label{ncatm1} \end{eqnarray}
where $F\star G = F\,
\mbox{exp}\(\frac{\theta}{2}(\overleftarrow{\pa_{+}
}\overrightarrow{\pa_{-}}-\overleftarrow{\pa_{-}}
\overrightarrow{\pa_{+}})\) G $,\,\, and  $g \in U(1)\mbox{x}
U(1)$. $I_{WZW}[g]$ is the NC generalization of a WZNW action for
$g$

\br I_{WZW}[g] =  \int d^2x \left[ \pa_{+}  g \star \pa_{-} g^{-1}
+ \int_0^1 dy \hat{g}^{-1} \star \pa_{y} \hat{g} \star
\[\hat{g}^{-1} \star \pa_{+}  \hat{g}, \hat{g}^{-1} \star \pa_{-}
\hat{g}
\]_{\star}\right], \er where the homotopy path $\hat{g}(y)$ such
that $\hat{g}(0)=1$, $\hat{g}(1) = g$ ($[y,x_{+}]=[y,x_{-}]=0$)
has been defined. The WZW term in this case gives a non-vanishing
contribution due to the noncommutativity. This is in contrast with
the action (\ref{atmmat}) in ordinary space. Notice that we have
introduced two independent fields (one real field for each $U(1)$
group) instead of the complex field $\vp$. This is justified since
in the NC realm the Abelian subgroup of $GL(2)$ fails to decouple
from the rest of the fields of the model as we will show below.

From (\ref{ncatm1}) one can derive the  set of equations of motion
for the corresponding fields \br
  \label{eqmg1}
 \pa_{-}(g^{-1}\star \pa_{+} g)& =& \[F^{-}\,,\, g\star F^{+}\star g^{-1}\]_{\star}
\\
\label{eqmf1}
\partial_{+} F^{-} &=& [E_{-2}, \partial_{+} W^{+}
],\,\,\,\,\,\,\,\,\,
\partial_{-} F^{+} = -[E_{2}, \partial_{-} W^{-} ],
 \\
 \partial_{+} W^{+} &=& - g \star F^{+} \star g^{-1},\,\,\,\,
  \partial_{-} W^{-} = - g^{-1} \star F^{-} \star g \label{eqmw1}.
                \end{eqnarray}
Substituting the derivatives of $W^{\pm}$'s given in the Eqs.
(\ref{eqmw1}) into the Eqs. (\ref{eqmf1}) one can get the
equivalent set of equations \br \label{eqmf11}
\partial_{+} F^{-}& =& -[E_{-2}\,,\,  g \star F^{+} \star g^{-1} ],
\,\,\,\,\,\,\,\,\,\,\,
\partial_{-} F^{+} = [E_{2}\,,\,  g^{-1} \star F^{-} \star g ].
\er

Notice that in the action (\ref{ncatm1}) one can use
simultaneously the cyclic properties of the group trace and the
$\star$ product. Then, the action (\ref{ncatm1}) and the equations
of motion (\ref{eqmg1})-(\ref{eqmw1}) have the left-right local
symmetries given by \br \label{sym1}
g & \rightarrow & h_{L}(x_{-}) \star g(x_{+}, x_{-}) \star  h_{R}(x_{+}),\\
\label{sym2} F^{+} &\rightarrow& h^{-1}_{R}(x_{+})\star
F^{+}(x_{+}, x_{-}) \star  h_{R}(x_{+}),\,\,\,\, W^{-} \rightarrow
h^{-1}_{R}(x_{+})\star W^{-}(x_{+}, x_{-}) \star
h_{R}(x_{+}),\,\,\,\,\,\,
\\
\label{sym3} F^{-} &\rightarrow& h_{L}(x_{-})\star F^{-}(x_{+},
x_{-})\star h^{-1}_{L}(x_{-}),\,\,\,\,W^{+} \rightarrow
h_{L}(x_{-})\star W^{+}(x_{+}, x_{-})\star  h^{-1}_{L}(x_{-}).\er

 In fact, the system of Eqs. (\ref{eqmg1})-(\ref{eqmw1}) is invariant under the above symmetries if the following conditions are supplied
\br \label{condi} h_{R}(x_{+})\,\star E_{2}\,
h^{-1}_{R}(x_{+})\,=\, E_{2},\,\,\,\,\, h^{-1}_{L}(x_{-})\,\star
E_{-2}\,  h_{L}(x_{-})\,=\, E_{-2}, \er where $h_{L/R}(x_{\mp})\,
\in\, {\cal H}_{0}^{L/R}$, \, ${\cal H}_{0}^{L/R}$ being Abelian
sub-groups of $GL(2, C)$.

Next, we define the {\sl second} version of the NC affine Toda
model coupled to matter NCATM$_{2}$ as \br \label{ncatm2}
S_{NCATM_{2}} \equiv S[g, W^{\pm},F^{\pm}] + S[\bar{g}, {\cal
W}^{\pm},{\cal F}^{\pm}], \er where the independent fields $g$\,
and\, $\bar{g}$, related to the set of matter fields $\{W^{\pm},
F^{\pm}\}$ and $\{ {\cal W}^{\pm},{\cal F}^{\pm}\}$, respectively,
belong to complexified \,$U(1)_{C}$ \, groups with the action
$S[.\, , \, . \, , \, .]$ being a Moyal extension of
(\ref{atmmat}) for $g \in U(1)_{C}$.

The equations of motion for the NCATM$_{2}$ model (\ref{ncatm2} )
comprise the Eqs. (\ref{eqmg1})-(\ref{eqmw1}) written for $g \in
U(1)_{C}$ and a set of analogous equations for the remaining
fields $\bar{g}$, ${\cal F}^{\pm}$ and ${\cal W}^{\pm}$. Moreover,
in addition to the symmetry transformations
(\ref{sym1})-(\ref{sym3}) one must consider similar expressions
for $\bar{g}$, ${\cal F}^{\pm}$ and ${\cal W}^{\pm}$.

\section{NC versions of the sine-Gordon model (NCSG$_{1,2}$)}

\label{ncsg12}

In order to derive the NC versions of the sine-Gordon model we
follow the master Lagrangian approach \cite{deser}, as in the
ordinary SG derivation \cite{jhep}, starting from the
NCATM$_{1,2}$ models (\ref{ncatm1}) and  (\ref{ncatm2}),
respectively. Let us concentrate first on the equations of motion
(\ref{eqmg1})-(\ref{eqmw1}) which are understood to be written for
$g \in U(1)\mbox{x} U(1)$ or $U(1)_{C}$. We proceed by considering
the Eqs. (\ref{eqmf1}) and integrating them
\begin{eqnarray}
 \label{solf1}
 F^{-} = [E_{-2},  {W}^{+} ] + f^{-}(x_{-}),\,\,\,\,
 F^{+} =- [E_{2},  {W}^{-} ] - f^{+}(x_{+}).
        \end{eqnarray}
with the $f^{\pm}(x_{\pm})$'s being analytic functions. Next, we
replace the $F^{\pm}$ of Eqs. (\ref{solf1}) and the $\pa_{\pm}
W^{\pm}$ of (\ref{eqmw1}), written in terms of $W^{\pm}$, into the
action (\ref{ncatm1}) to get
                \begin{eqnarray}
    \nonumber            S'[g, W^{\pm}, f^{\pm}] &=& I_{WZW}[g] + \int d^2x \{ \frac{1}{2}< [E_{-2},W^{+}] \star g
                \star f^{+} \star g^{-1}> + \\
&& \frac{1}{2}< g^{-1}\star f^{-}
                \star g \star [E_{2}, W^{-}] > +
                <g^{-1}\star f^{-} \star g \star f^{+}> \}\label{ncsg1aux}
                \end{eqnarray}

As the next step, one writes the equations of motion for the
$f^{\pm}(x_{\pm})$'s and solves for  them; afterwards, substitutes
those expressions into the intermediate action (\ref{ncsg1aux})
getting
 \begin{eqnarray}
                S''[g, W^{\pm}] =  I_{WZW} [g] - \frac{1}{4}\int d^2x[ <[E_{-2},W^{+}] \star g \star
                [E_{2},W^{-}] \star g^{-1}>]
\label{ncsgeff}.                \end{eqnarray}

Notice that (\ref{ncsgeff}) has inherited from the NCATM action
the local symmetries (\ref{sym1})-(\ref{sym3}). Therefore, one
considers the gauge fixing
                \begin{eqnarray}
\label{gf}
                2i \Lambda^{-} = [E_{-2},W^{+}], \;\;\;
                2i \Lambda^{+}  =[E_{2},W^{-}],
                \end{eqnarray}
where $\Lambda^{\pm} \in \hat{{\cal G}}_{\pm 1}$ are some constant
generators in the subsets of grade $\pm 1$ (\ref{eigensl2}).

 Then for this gauge fixing the effective action (\ref{ncsgeff}) becomes
                \begin{eqnarray}\nonumber
                S_{NCSG_{1}}[g] &\equiv & S[g] \\
                &= & I_{WZW} [g] +\int d^2x [ < \Lambda^{-}  \star g \star \Lambda^{+}
                \star g^{-1}>].  \label{ncsg1}
                \end{eqnarray}

Then, the equation of motion for the field $g$ is \br
\label{eqncsg1}
 \pa_{-}(g^{-1}\star \pa_{+} g)& =&\[\L^{-}\,,\, g \star \L^{+} g^{-1}\]
\er

The action (\ref{ncsg1}) for $g\in U(1) \mbox{x} U(1)$ defines the
{\sl first} version of the non-commutative sine-Gordon model
(NCSG$_{1}$) (see Sec. \ref{lech}).

Since the above reduction process can be carried over verbatim for
each sector of the NCATM$_{2}$ model (\ref{ncatm2}) and its
independent set of fields $\{ g, F^{\pm}, W^{\pm}\}$ and $\{
\bar{g}, {\cal F}^{\pm}, {\cal W}^{\pm}\}$\,, one can write in
this case as the reduced model \br \label{ncsg2} S_{NCSG_{2}}[g,
\bar{g}]= S[g] + S[\bar{g}];\,\,\,\,\,\,\,\, g,\, \bar{g} \in
U(1)_{C}. \er

The equations of motion derived from this action become the Eq.
(\ref{eqncsg1}) written for $g \in U(1)_{C}$ and  \br
  \label{eqncsg2}
 \pa_{-}(\bar{g}^{-1}\star \pa_{+} \bar{g})& =&\[\L^{-}\,,\, \bar{g} \star \L^{+} \bar{g}^{-1}\]
.\er

The action (\ref{ncsg2}) defines the {\sl second} version of the
non-commutative sine-Gordon model (NCSG$_{2}$) (see Sec.
\ref{GP}).

In the subsections below we will see that NCSG$_{1}$ (\ref{ncsg1})
written for $g\in U(1) \mbox{x} U(1)$ and NCSG$_{2}$ (\ref{ncsg2})
are precisely the Lechtenfeld et al. \cite{lechtenfeld} and
Grisaru-Penati \cite{grisaru2} proposals for the NC versions of
the SG model, respectively.

\subsection{Lechtenfeld et al. proposal (NCSG$_{1}$)}

\label{lech}

The NCSG$_{1}$ version has been obtained through the reduction
process starting from the NCATM$_{1}$ model (\ref{ncatm1}), so let
us write the field $g \in U(1)$x$U(1)$ in the representation \br
\label{u1u1} g = \left(\begin{array}{cr}
e^{i\vp_{+}}_{\star} &  0 \\
0 & e^{-i\vp_{-}}_{\star}
\end{array} \right)\,\equiv\, g_{+} g_{-},\,\,\,\,
g_{+} = \left(\begin{array}{cr}
e^{i\vp_{+}}_{\star} &  0 \\
0 & 1
\end{array} \right),\,\,\,\, g_{-} = \left(\begin{array}{cr}
1 &  0 \\
0 & e^{-i\vp_{-}}_{\star}
\end{array} \right)
\er with $\vp_{\pm}$ being real fields.

For the $\Lambda$'s taken as
   \br \label{consts} \Lambda^{+}=
   M(E_{+}^{0}+E_{-}^{1}),\,\,\,\Lambda^{-}= M(E_{-}^{0}+E_{+}^{-1}),
   \er
the action (\ref{ncsg1}) for $g$ given in  (\ref{u1u1}), upon
using the Polyakov-Wiegmann identity, can be written as
\footnote{Assuming the general forms $\L^{+} =\( \L_R\,
                E_+^0 + \widetilde \L_R E_-^1\), \,  \L^{-} =\(
\L_L\,
                E_+^{-1} +  \widetilde \L_L\, E_-^0 \),\,
                 $ one  gets
               $S[g]=  I_{WZW} [g] +  \int [\L_L
\widetilde{\L}_R
               e^{-i \vp_{-}}_{\star} \star e^{-i\vp_{+}}_{\star}+\widetilde{\L}_L \L_R
               e^{i \vp_{+}}_{\star} \star e^{i\vp_{-}}_{\star} ]$, which upon setting
$\widetilde{\L}_L
                \L_R=2e^{i\d}M^2, \, (\d = 0)$ reproduces (\ref{nclech11})
(the
                phase $\d \neq 0$ can be absorbed by shifting the
                fields
$\vp_{\pm}$).}
\begin{eqnarray}\label{nclech11}
                   S_{NCSG_{1}}[g_{+}, g_{-}] = I_{WZW}[g_{+}] +
                   I_{WZW}[g_{-}] + M^{2} \int d^2x
                   \( e_{*}^{i \vp_{+}}\star e^{ i \vp_{-}}_{\star}+ e_{*}^{-i \vp_{-}} \star e^{- i \vp_{+}}_{\star}-2
                   \).
   \end{eqnarray}

In this way we have re-derived the Lechtenfeld et al. action
(NCSG$_{1}$) for the NC sine-Gordon \cite{lechtenfeld}. The Eqs.
of motion become \br\label{ncsgeq1}
\pa_{-}\(e_{\star}^{-i\vp_{+}}\star \pa_{+} e_{\star}^{i
\vp_{+}}\)&=& - M^{2} \(e^{i\vp_{+}}_{\star} \star e_{\star}^{i
\vp_{-}}- e^{-i\vp_{-}}_{\star} \star e_{\star}^{-i \vp_{+}}\);\\
\pa_{-}\(e_{\star}^{i\vp_{-}} \star \pa_{-} e_{\star}^{-i \vp_{-}}
\)&=& +M^{2} \(e^{i\vp_{+}}_{\star} \star e_{\star}^{i \vp_{-}}-
e^{-i\vp_{-}}_{\star} \star e_{\star}^{-i
\vp_{+}}\).\label{ncsgeq2}\er

In the $\theta \rightarrow 0$ limit the above equations can be
written as \br \label{limit1}\pa_{-}\pa_{+} \(\vp_{+}-\vp_{-}\)&=&0;\\
\pa_{-}\pa_{+}\(\vp_{+}+\vp_{-}\)&=& -4M^{2}
\mbox{sin}\(\vp_{+}+\vp_{-}\).\label{limit2}\er

If we choose $\vp_{+}=\vp_{-} \equiv \frac{1}{2}
\vp_{SG}\leftrightarrow e^{i\vp_{+}}=e^{i\vp_{-}} \in U(1)_{A}$,
we have in (\ref{limit2}) the SG equation $\pa^{2} \vp_{SG} = -4
M^2 \mbox{sin} (\vp_{SG})$. Thus the $U(1)_{V}$ degree of freedom
completely decouples in the commutative limit. Then, in the model
NCSG$_{1}$ one can define the topological charge as\br
\label{chargencsg1} Q_{\rm topol.}^{NCSG_{1}}= \frac{1}{2\pi}
\int_{-\infty}^{+\infty} dx \frac{d \(\vp_{+}+\vp_{-}\)}{dx}
\equiv \sum_{n} \theta^n Q^{(n)}_{NCSG_{1}}. \er

The Leznov-Saveliev  formulation of the NCSG$_{1}$ model through a
zero-curvature equation will lead to (\ref{eqncsg1}) for the
parametrization (\ref{u1u1}). A linear system for this system is
provided in \cite{lechtenfeld} through a dimensional reduction
from $(2+2)$ self-dual Yang Mills theory. Then, following a
similar procedure to that developed in \cite{cabrera} (see also
\cite{Hamanaka1}) for a NC linear system one can construct
infinite conserved currents.

\subsection{The Grisaru-Penati proposal (NCSG$_{2}$)}

\label{GP}

        The second NC version of the sine-Gordon system (NCSG$_{2}$)
         is written in terms of the following
        representation of the complexified $U(1)_{C}$ group elements
                   \br \label{exp}
                           g= e^{i\vp H^{0}}_{\star} \equiv
                           \left(\begin{array}{cr}
                           e^{i\vp}_{\star} &  0 \\
                           0 & e^{-i\vp}_{\star}
                           \end{array} \right)\,\,\,\,\, \mbox{and}\,\,\,\,\,
                                                                                             \bar{g}= e^{i(\vp)^{\dagger} H^{0}}_{\star} \equiv
 \left(\begin{array}{cr}
e^{i\vp^{\dagger}}_{\star} &  0 \\
 0 & e^{-i\vp^{\dagger}}_{\star}\end{array} \right), \er
where the field $\vp$ is a general complex field and the $g$ and
 $\bar{g}$ are formally considered as independent fields.

 The master Lagrangian from which the NCSG$_{2}$ model
    originates is the NCATM$_{2}$ theory (\ref{ncatm2}). Thus, one must
     consider the reduced model (\ref{ncsg2}).

Thus for the $\Lambda$'s given in (\ref{consts}) the action
(\ref{ncsg2}), taking into account the action (\ref{ncsg1})
written for $g \in U(1)_{C}$, can be written as
\begin{eqnarray}\label{ncsg21}
                S_{NCSG_{2}} &=& I_{WZW}[g] +  M^{2} \int d^2x
                \mbox{Tr} \( g^{2} +g^{-2}-2 \) +\\ && I_{WZW}[\bar{g}] +  M^{2} \int d^2x \mbox{Tr} \( \bar{g}^{2} +\bar{g}^{-2}-2 \)
\end{eqnarray}
where $I_{WZW}[g]$ is the NC generalization of a complexified
$U(1)$ WZNW action \cite{naculich}.

In this way we have arrived at the Grisaru-Penati proposal for the
 NC sine-Gordon system (NCSG$_{2}$)\cite{grisaru1, grisaru2}.

Notice that, when the field $\vp$ is real, one has $g=\bar{g}$ and
the action reduces to $S_{NCSG_{2}}[g, \bar{g}]=2\[ I_{WZW}[g] +
\int M^{2}
                \mbox{Tr} \( g^{2} +g^{-2}-2 \)\] $. In fact, it is possible to find real solutions for the
                NCSG$_{2}$ model \cite{grisaru1}.

Regarding the NCSG$_{1,2}$ relationships, notice that assuming
general complex fields $\vp_{\pm}$ and imposing the reduction
$\vp_{+}=\vp_{-}\equiv \vp$ in the NCSG$_{1}$ action
(\ref{nclech11}) one may get the $S[g]$ sector of the NCSG$_{2}$
model (\ref{ncsg21}).

The Leznov-Saveliev  formulation of the NCSG$_{2}$ model
 \cite{cabrera} through  a zero-curvature equation leads to Eqs.
 (\ref{eqncsg1}) and
 (\ref{eqncsg2}).

\section{Decoupling of  NCSG$_{1,2}$ and NCMT$_{1,2}$ models}

\label{decou}

In the study of the ordinary (commutative) ATM model performed in
\cite{nucl}-\cite{jhep}, the massive Thirring model (MT) was
obtained by means of a Hamiltonian reduction and the so-called
decoupling procedures. The first procedure requires the definition
of conjugated momenta for the fields of the model. In the NC case
this procedure encounters some complications due to the infinite
sum of time derivatives implicit in the Moyal product (see, e.g.
\cite{barcelos}) and then we must resort to an alternative method
to uncover the MT sector of the NCATM theories. In \cite{nucl,
jhep} it has been proposed another approach to recover the SG and
MT models out of  the ordinary ATM model. This proceeds by
decoupling the set of equations of the ATM model into the
corresponding dual models. This procedure can be adopted in the NC
case by writing a set of mappings between the fields of the model
such that the Eqs. (\ref{eqmg1}) and (\ref{eqmf11}) when rewritten
using those mappings completely decouple the scalar and the matter
fields. Following the commutative case let us consider the
mappings \br \label{decg1}
 \[F^{-}\,,\, g\star F^{+}\star g^{-1}\]_{\star}& =& \[\L^{-}\,,\, g\star \L^{+}\star g^{-1}\]
\\
\label{decf1}
-\[E_{-2}, g\star F^{+}\star g^{-1}\] &=& \[E_{-2},\[E_{2}, W^{-}\]\] - \nonumber \\
&& \frac{\l}{8} \[\[E_{-2}, W^{+}\], \[\[ E_{2},
W^{-}\],W^{-}\]_{\star}\]_{\star}\\
\label{decf2}
\[E_{2}, g^{-1}\star F^{-}\star g\] &=& \[E_{2},\[E_{-2},
W^{+}\]\] - \nonumber  \\
&& \frac{\l}{8} \[\[E_{2}, W^{-}\], \[\[ E_{-2},W^{+}\],
W^{+}\]_{\star}\]_{\star}
\\
\label{decef} F^{\pm}& =  & \mp [E_{\pm 2}\,,\,W^{\mp}]. \er

In the relations  above the field $g$ comes from section 2 and we
assume it belongs to either $U(1)\mbox{x}U(1)$ or $U(1)_{C}$. In
writing the mappings (\ref{decg1})-(\ref{decef}) a helpful
organizing guide is the principal gradation structure (see
Appendix \ref{appa}) such that only equal grade terms ($0$\, or \,
$\pm 1$)\, appear in each relationship.

It is clear that the NCSG$_{1, 2}$ (one sector of model $2$)
 equation of motion (\ref{eqncsg1}) is recovered from
 the equation of motion (\ref{eqmg1}) and the decoupling
 equation (\ref{decg1}). We expect that a noncommutative version of the massive
Thirring model (NCMT$_{1}$) defined for the fields $W^{\pm}$,
corresponding to the Letchenfeld et al. version NCSG$_{1}$, will
emerge from the decoupling Eqs. (\ref{decf1})-(\ref{decef}) and
the Eqs. of motion (\ref{eqmf11}).

In order to recover the Grisaru-Penati version  NCSG$_{2}$ one
must
 write similar decoupling expressions for the full set of fields $\{g, F^{\pm}, W^{\pm} \}$\, and\, $\{\bar{g}, {\cal F}^{\pm}, {\cal W}^{\pm}\}$. Thus,
 following similar steps to the previous construction
  we expect to recover another version of the NC massive Thirring model NCMT$_{2}$ defined for the
  fields $\{W^{\pm}, {\cal W}^{\pm}\}$. In the next section we
  propose two versions of the non-commutative massive Thirring theories (NCMT$_{1,2}$) by
  providing the Lagrangians and the zero-curvature equations.

\section{The NC (Bosonic) Thirring models (B)NCMT$_{1, 2}$}

\label{ncmt12}

In ordinary space the formulation of the MT model can be performed
in two ways. First, the classical fields can be assumed to be
anti-commuting Grassmannian fields \cite{izergin}. Second, the
fields $\psi, \widetilde{\psi}$ considered as ordinary commuting
fields define the so-called bosonic massive Thirring (BMT)
\cite{orfanidis, garbaczewski}. Even though in \cite{annals, jhep}
the authors have been considered anti-commuting fields in order to
make the reduction procedure of the relevant ATM models into its
dual theories, here, we follow the second formulation, i.e. we
will consider commuting fields. The reason is that the
zero-curvature formulations of the NCMT$_{1,2}$ models follow from
that of the ATM relevant formulation in the context of the affine
Lie algebra $sl(2)$ construction. This point of view is also in
accordance with the assumption in \cite{matter} where these fields
have been considered as ordinary commuting fields leaving the
discussion of their fermionic character to the full quantum
treatment of the models, since the statistics of fields in
two-dimensions depends upon the coupling constant. Regarding this
point, it is known that already in the (B)MT classical solutions
(see, e.g. \cite{orfanidis}) it has been discussed the appearance
of certain Pauli exclusion principle associated with the
multi-soliton solutions in the context of the classical
correspondence between the SG and the MT models.

The decoupling procedure of the NCAMT$_{1, 2}$ models provide two
models which we shall call (bosonic) non-commutative massive
Thirring models [(B)NCMT]$_{1, 2}$, respectively, in the
derivations below.

\subsection{(B)NCMT$_{1}$}

\label{bncmt1}

We propose the (B)NCMT$_{1}$ Lagrangian related to the fields
$W^{\pm}$ such that it reproduces the relevant equations of motion
we have outlined in the last section by the decoupling procedure
of the NCATM model. Let us consider the action

\begin{eqnarray} \label{ncmt1}
S[W^{\pm},\widetilde{W}^{\pm} ] &=& \int \{ <[ E_{-2},
\widetilde{W}^{+}] \star \partial_{+} W^{+}> - < \partial_{-}
W^{-} \star [
E_{2},\widetilde{W}^{-}]>\nonumber \\
& & -< [E_{-2},\widetilde{W}^{+}] \star [E_{2},W^{-}]> -<
[E_{-2},W^{+}] \star
[E_{2}, \widetilde{W}^{-}] > - \nonumber  \\
& & \l <J^{-} \star J^{+}> \}
\end{eqnarray}
where the current components are given by
\begin{eqnarray}
\label{curr1}
 J^{+} &=& \frac{1}{4} \( [[E_{-2}, \widetilde{W}^{+}],
W^{+}]_{\star}+[[E_{-2}, W^{+}],
\widetilde{W}^{+}]_{\star}\) \\
\label{curr2} J^{-} &=&- \frac{1}{4} \([[E_{2}, W^{-}],
\widetilde{W}^{-}]_{\star} + [[E_{2}, \widetilde{W}^{-}],
W^{-}]_{\star}\).
\end{eqnarray}

In order to write the expressions in more symmetric form we have
considered additional fields denoted by $\widetilde{W}^{\pm}$ (see
below). We will show that the action (\ref{ncmt1}) is related to
the  NCMT$_{1}$ version. In the derivations below we use the
explicit matrix representation (\ref{basis}) for the $GL(2)$
generators and its corresponding loop extension. The field
components are defined by \br \label{spinors1}
W^{+}&=&\sqrt{\frac{4i}{m_{\psi}}}\(\psi_{L} E_{+}^{0}+
\widetilde{\psi}_{L} E_{-}^{1}
\),\,\,\,\,\,\,\,\,\,\,\,\,\,\,\,\,\,\,\,\,\,\,
W^{-}=-\sqrt{\frac{4i}{m_{\psi}}}\(\psi_{R} E_{+}^{-1}- \widetilde{\psi}_{R} E_{-}^{0} \)\\
\[E_{-2}\,,\,\widetilde{W}^{+}\]&=&-\sqrt{\frac{im_{\psi}}{4}}\(\psi_{L}
E_{+}^{-1}- \widetilde{\psi}_{L} E_{-}^{0} \),\,\,\,\,
\[E_{2}\,,\,\widetilde{W}^{-}\]=\sqrt{\frac{i m_{\psi}}{4}}\(\psi_{R}
E_{+}^{0}+\widetilde{\psi}_{R} E_{-}^{1} \). \label{spinors2} \er

Then the (B)NCMT$_{1}$ action (\ref{ncmt1}) in terms of the field
components is given by \br {\cal S}_{(B)NCMT_{1}}&=& \int d^2x \[
 \nonumber
 2i\widetilde{\psi}_{L}\pa_{+}\psi_{L} +
 2i\widetilde{\psi}_{R}\pa_{-}\psi_{R}-im_{\psi}\(\widetilde{\psi}_{R}\psi_{L}
-\widetilde{\psi}_{L}\psi_{R}\)-\\
&&\l\(\widetilde{\psi}_{R}\star \psi_{R}\star
\widetilde{\psi}_{L}\star \psi_{L}+ \psi_{R}\star
\widetilde{\psi}_{R}\star \psi_{L}\star \widetilde{\psi}_{L}
\)\]\label{ncmtfield} \er

Notice that although $W^{\pm}$ and $\widetilde{W}^{\pm}$ are
proportional, we consider them as independent fields since the
waved fields appear inside the Lie bracket making the expressions
$[E_{\pm 2}\,,\, \widetilde{W}^{\mp}]$ indeed independent from the
fields $W^{\pm}$. This notation will be useful in order to derive
the field equations in matrix form starting from the action
(\ref{ncmt1}).

By taking the functional derivative of (\ref{ncmt1}) with respect
to $\widetilde{W}^{+}$, $\widetilde{W}^{-}$, $W^{+}$ and $W^{-}$,
respectively, one can get the equations of motion \br
\[E_{-2}\,,\,\pa_{+}W^{+} \] &=& \[E_{-2},\[E_{2},
W^{-}\]\] +  \frac{\l}{8}\, \[\[E_{-2}, W^{+}\],
\[\[E_{2},\widetilde{W}^{-}\], W^{-}\]_{\star}+\nonumber
\\
&& \[\[E_{2},W^{-}\], \widetilde{W}^{-}\]_{\star}\]_{\star}
\label{eqncmt1}
\\
 \[E_{2}\,,\,\pa_{-}W^{-} \]&=&- \[E_{2},\[E_{-2},
W^{+}]\] -  \frac{\l}{8}\, \[\[E_{2}, W^{-}\],
\[\[E_{-2},\widetilde{W}^{+}\], W^{+}\]_{\star}+\nonumber\\
&&\[\[E_{-2},W^{+}\], \widetilde{W}^{+}\]_{\star}\]_{\star}.
\label{eqncmt2}
\\
\[E_{-2}\,,\,\pa_{+}\widetilde{W}^{+} \]  &=& \[E_{-2},\[E_{2},
\widetilde{W}^{-}\]\] + \frac{\l}{8}\, \[\[E_{-2},
\widetilde{W}^{+}\], \[\[
E_{2},W^{-}\],\widetilde{W}^{-}\]_{\star} +
\nonumber\\
&&\[\[ E_{2},\widetilde{W}^{-}\],W^{-}\]_{\star} \]_{\star}
\label{eqncmt3}
\\
\[E_{2}\,,\,\pa_{-} \widetilde{W}^{-} \] &=&  - \[E_{2},\[E_{-2}, \widetilde{W}^{+}\]\] -
\frac{\l}{8}\, \[\[E_{2},
\widetilde{W}^{-}\]\,,\,\[\[E_{-2},W^{+}\],
\widetilde{W}^{+}\]_{\star} + \nonumber
\\
&&\[\[E_{-2},\widetilde{W}^{+}\], W^{+}\]_{\star}
\]_{\star}\label{eqncmt4} \er

The Eqs. (\ref{eqncmt3})-(\ref{eqncmt4}) when considered in terms
of the field components (\ref{spinors1})-(\ref{spinors2}) are
corresponding  copies of the Eqs. (\ref{eqncmt1})-(\ref{eqncmt2}),
so in the considerations  below it will be sufficient to pay
attention only on these equations. One can verify that the
equations of motion (\ref{eqncmt1})-(\ref{eqncmt2}) reproduce the
set of equations obtained when the decoupling mappings
(\ref{decf1})-(\ref{decef}) are replaced into the equations
(\ref{eqmf11}).

The equations of motion (\ref{eqncmt1})-(\ref{eqncmt2}) for the
fields defined by (\ref{spinors1})-(\ref{spinors2}) become \br
\label{ncth1} \pa_{+} \psi_{L} &=& - \frac{m_{\psi}}{2}\psi_{R}-\,
i\, \frac{\l}{2} \, \(\psi_{L}\star \widetilde{\psi}_{R}\star
\psi_{R}+\psi_{R}\star \widetilde{\psi}_{R} \star \psi_{L}\),\\
\label{ncth2} \pa_{-} \psi_{R}  &=&\,\,\,\,
\frac{m_{\psi}}{2}\psi_{L}-\, i\, \frac{\l}{2} \, \(\psi_{R}\star
\widetilde{\psi}_{L}\star \psi_{L} +\psi_{L}\star
\widetilde{\psi}_{L} \star \psi_{R} \)\\
\label{ncth3} \pa_{+} \widetilde{\psi}_{L} &=& -
\frac{m_{\psi}}{2}\widetilde{\psi}_{R}+\, i\, \frac{\l}{2} \,
\(\widetilde{\psi}_{L}\star \psi_{R}\star
\widetilde{\psi}_{R}+\widetilde{\psi}_{R}\star \psi_{R} \star \widetilde{\psi}_{L}\),\\
\label{ncth4} \pa_{-} \widetilde{\psi}_{R}  &=&\,\,\,\,
\frac{m_{\psi}}{2}\widetilde{\psi}_{L}+\, i\, \frac{\l}{2} \,
\(\widetilde{\psi}_{R}\star \psi_{L}\star \widetilde{\psi}_{L}
+\widetilde{\psi}_{L}\star \psi_{L} \star \widetilde{\psi}_{R} \).
\er

Notice that in the limit $\theta \rightarrow 0$ the equations
(\ref{ncth1})-(\ref{ncth4}) reduce to the usual (B)MT Eqs. of
motion for $\widetilde{\psi}_{R,\, L}=\psi^{\star}_{R,\, L}$
($\star$ here means complex conjugation) \cite{orfanidis,
garbaczewski}.

The system of Eqs. (\ref{eqncmt1})-(\ref{eqncmt2}) admit a
zero-curvature formulation.  In fact, consider \br \label{conn1}
A_{-}&=&E_{-2} + i \sqrt{\frac{\l}{4}}\,\,[E_{-2}\,,\,W^{+} ] +\frac{\l}{4}\,\, [[E_{-2}\,,\,\widetilde{W}^{+} ]\,,\,W^{+}]_{\star},\\
A_{+}&=&-E_{2} -i  \sqrt{\frac{\l}{4}}\, \,[E_{2}\,,\,W^{-} ]
-\frac{\l}{4}\,\, [[E_{2}\,,\,\widetilde{W}^{-}
]\,,\,W^{-}]_{\star}. \label{conn2} \er

Then from the zero-curvature condition\,  $[\pa_{+}+ A_{+}\, ,
\,\pa_{-}+ A_{-}\ ]_{\star}=0$\, one obtains the set of Eqs.
(\ref{eqncmt1})-(\ref{eqncmt2}) plus an additional equation \br &&
\nonumber \pa_{+}[[E_{-2}\,,\,\widetilde{W}^{+}
]\,,\,W^{+}]_{\star}+
\pa_{-}[[E_{2}\,,\,\widetilde{W}^{-} ]\,,\,W^{-}]_{\star}=\\
&&\[ [E_{2}\,,\,W^{-} ]\,,\,[E_{-2}\,,\,W^{+} ]\]-\frac{\l}{4} \[
[[E_{2}\,,\,\widetilde{W}^{-}
]\,,\,W^{-}]_{\star}\,,\,[[E_{-2}\,,\,\widetilde{W}^{+}
]\,,\,W^{+}]_{\star}\]. \label{additional1}
 \er

The Eq. (\ref{additional1}) in terms of the component fields gives
rise to the equation \br \nonumber \pa_{-}
(\widetilde{\psi}_{R}\star \psi_{R})-\pa_{+}
(\widetilde{\psi}_{L}\star \psi_{L})&=&  m_{\psi}
(\widetilde{\psi}_{R}\star \psi_{L}+\widetilde{\psi}_{L}\star
\psi_{R})-\\&& i\, \l (\widetilde{\psi}_{R}\star \psi_{R}\star
\widetilde{\psi}_{L}\star \psi_{L} - \widetilde{\psi}_{L}\star
\psi_{L}\star \widetilde{\psi}_{R}\star \psi_{R}) \label{adeq1}
 \er
and another equation obtained from (\ref{adeq1}) by conveniently
substituting \br \label{conjugation} \{\psi_{R},\,\psi_{L}
\}\leftrightarrow \{\widetilde{\psi}_{R},\,\widetilde{\psi}_{L} \}
\,\,\,\,\mbox{and}\,\,\,\, i \rightarrow -i .\er

The set of Eqs. (\ref{adeq1}) and the one obtained by
 (\ref{conjugation}) may be shown to be satisfied as the result of
the field equations (\ref{ncth1})-(\ref{ncth4}). The Eq.
(\ref{adeq1}) can be written as $\pa_{\mu} j^{(1)\,\mu}_{5} =2i
m_{\psi} \bar{\psi} \g_{5}\star \psi -2i\, \l
(\widetilde{\psi}_{R}\star \psi_{R}\star \widetilde{\psi}_{L}\star
\psi_{L} - \widetilde{\psi}_{L}\star \psi_{L}\star
\widetilde{\psi}_{R}\star \psi_{R})$, where $j^{(1)\,\mu}_{5}
\equiv \bar{\psi} \g^{\mu} \g_{5} \star \psi$. In the $\theta
\rightarrow 0$ limit the last term inside parenthesis in
(\ref{adeq1}) vanishes and the $j^{(1)\,\mu}_{5}$ current is
conserved for $m_{\psi}=0$. The expression for
$\pa_{\mu}j_{5}^{(2)\,\mu}$
 may be obtained by making the substitutions (\ref{conjugation}) in
the relevant terms of $\pa_{\mu}j_{5}^{(1)\,\mu}$.

The currents (\ref{curr1})-(\ref{curr2}) satisfy\,  $ \pa_{+}
J^{+}+\pa_{-}J^{-} = 0$\, or equivalently written in field
components \br \label{adeq2} \pa_{-} (\widetilde{\psi}_{R}\star
\psi_{R})+\pa_{+} (\widetilde{\psi}_{L}\star \psi_{L})&=&0
;\,\,\,\,\,\pa_{-} (\psi_{R}\star \widetilde{\psi}_{R})+\pa_{+}
(\psi_{L}\star \widetilde{\psi}_{L})\,=\,0 \er

The (B)NCMT$_{1}$ model in (\ref{ncmt1}) [or (\ref{ncmtfield})]
has a global $U(1)$ symmetry. In order to obtain the currents by
the Noether procedure we make the global transformation localized,
as discussed in \cite{liao} this is not unique in the NC case. In
the equations (\ref{adeq2}) one recognizes the currents associated
to $U(1)\mbox{x} U(1)$ symmetry implemented in NC space with the
 transformation rules
\br \label{u1u1nc} \psi \rightarrow U_{2}(x) \star \psi \star
U_{1}^{-1}(x);\,\,\,\,\widetilde{\psi} \rightarrow U_{1}(x) \star
\widetilde{\psi} \star U_{2}^{-1}(x),
\,\,\,\,U_{1,2}(x)=e_{\star}^{i \a_{1,2}}, \er where $U_{1,2}(x)$
are independent starred exponentials with  $\a_{1,2}=\mbox{real
functions}$. In fact, the Eqs. (\ref{u1u1nc}) are the most general
transformations in NC space for a charged field \cite{liao}.

The $U_{1}(1)$ global symmetry of the action (\ref{ncmtfield})
gives through the Noether procedure the conservation equation
$\pa_{\mu} j^{(1)\,\mu}=0$;\, $j^{(1)\,\mu} \equiv \bar{\psi}\star
\g^{\mu} \psi$, where $\bar{\psi}= \widetilde{\psi}^{T} \g^{0}$,
corresponding to the first Eq. in (\ref{adeq2}). The another
$U_{2}(1)$ current conservation equation becomes $\pa_{\mu}
j^{(2)\,\mu}=0$;\,$j^{(2)\,\mu}\equiv -\psi^{T} \g^{0}\star
\g^{\mu} \widetilde{\psi}$ and corresponds to the second Eq. in
(\ref{adeq2}). Since the charge is associated to global
transformation of the charged field for which there is no
difference between the ordinary and non-commutative product one
can conclude that the currents share the same charge. In fact, for
global $U_{1, 2}$ only the product $U=U_{2} U^{-1}_{1}$ is
relevant. In this way we have uncovered the symmetry $U(1)\mbox{x}
U(1)$ \,in  (\ref{u1u1}) of the NCSG$_{1}$ model in the process of
constructing the conserved currents of the corresponding
NCMT$_{1}$ sector. Notice that the currents $j^{(1)\, \mu}$ and
$j^{(2)\, \mu}$ differ only by a sign in the commutative limit
(recall the bosonic nature of the matter fields); not so on NC
Euclidean space.

Moreover, a copy of the connection (\ref{conn1})-(\ref{conn2})
with the changes $\widetilde{W}^{\pm} \leftrightarrow W^{\pm}$
together with a corresponding zero-curvature equation reproduces
the other set of equations (\ref{eqncmt3})-(\ref{eqncmt4}).

Therefore, one can conclude that the Lechtenfeld et al. NCSG$_{1}$
model for two real fields ($\vp_{\pm}$) of section \ref{lech}
corresponds to the NCMT$_{1}$ theory defined in (\ref{ncmt1})  for
two types of matter fields $\widetilde{\psi}$ and $\psi$. In
section \ref{ncsol} we discuss this correspondence on the level of
the solitonic solutions of the NCATM$_{1}$ model.

In the NCSG$_{1}$ sector one must have a topological charge
corresponding to the above Noether charge.  The physical scalar
fields associated to this topological charge may be correctly
identified in the commutative limit $\theta \rightarrow 0$ of the
NCSG$_{1}$ Eqs. of motion (\ref{limit1})-(\ref{limit2}). In fact,
the combination $(\vp_{+}+ \vp_{-})$ carries the charge in the
NCSG$_{1}$ sector as defined in (\ref{chargencsg1}). This
correspondence can be better understood in the context of the
decoupling sectors of the NCATM$_{1}$ model such that the
one-soliton solution satisfies the Noether and topological
currents equivalence (\ref{equiv2}) also in the NC case (see
below).

Let us disclose some comments on the NCMT$_{1}$ action written for
Dirac fermions. The NCMT$_{1}$ Lagrangian (\ref{ncmtfield})
contains the non-standard interaction term $j^{(2)}_{\mu}\star
j^{(2)\,\mu} \sim
\psi_{R}\star\widetilde{\psi}_{R}\star\psi_{L}\star\widetilde{\psi}_{L}$,
which, to our knowledge, has not been considered previously in the
literature. The bosonization process of the NC extension of the
usual Thirring interaction performed in \cite{moreno, nunez}
considers only the interaction term $j^{(1)}_{\mu}\star
j^{(1)\,\mu}$. The NC version of the MT model with
$j^{(1)}_{\mu}\star j^{(1)\,\mu}$ interaction corresponds to a
bosonized theory whose dynamics is governed by a NC WZW action
plus a bosonized Thirring coupling and a NC cosine potential
\cite{nunez}, then the bosonized model resembles one of the
sectors, say $g$ sector, of the Grisaru-Penati model
(\ref{ncsg21}). It is expected that the bosonization procedure of
the NCMT$_{1}$ model will provide a bosonic action  of the
Lechtenfeld et al. type model (\ref{nclech11}). On the other hand,
gauge theories with fermions in the bi-fundamental representation
(\ref{u1u1nc}) in NC Euclidean space have been considered in
\cite{Nakajima} in order to study chiral anomalies in the NC
context.

\subsection{(B)NCMT$_{2}$}

As mentioned in the last paragraph of Section \ref{decou} we
expect that another NCMT$_{2}$ version will appear when one
performs similar decoupling processes for the extended system with
$\{F^{\pm}, W^{\pm}\}$ and $\{{\cal F}^{\pm}, {\cal W}^{\pm}\}$
fields. In fact, a copy of the  NCMT$_{1}$ action (\ref{ncmt1}),
as well as the relevant  zero-curvature equation of motion
 can be written for the fields $\{{\cal F}^{\pm}, {\cal W}^{\pm}\}$.

For ${\cal W}^{\pm}, \, \widetilde{{\cal W}}^{\pm}$  in components
we shall assume to contain the fields of type
$\({\Psi},\,\widetilde{\Psi}\)$ \br \label{spinors11} {\cal
W}^{+}\,=\,\sqrt{\frac{4i}{m_{\Psi}}}\(\Psi_{L} E_{+}^{0}+
\widetilde{\Psi}_{L} E_{-}^{1} \),\,\,\,\,\,\,\,\,\,\,
{\cal W}^{-}=-\sqrt{\frac{4i}{m_{\Psi}}}\(\Psi_{R} E_{+}^{-1}- \widetilde{\Psi}_{R} E_{-}^{0} \)\\
\[E_{-2}\,,\,\widetilde{{\cal W}}^{+}\]\,=\,-\sqrt{\frac{im_{\Psi}}{4}}\(\Psi_{L}
E_{+}^{-1}- \widetilde{\Psi}_{L} E_{-}^{0} \);\,\,
\[E_{2}\,,\,\widetilde{{\cal W}}^{-}\]=\sqrt{\frac{i m_{\Psi}}{4}}\(\Psi_{R}
E_{+}^{0}+ \widetilde{\Psi}_{R} E_{-}^{1} \). \label{spinors21}
\er

Thus, one can write the (B)NCMT$_{2}$ action for 4 types of matter
fields $\widetilde{\psi}\,, \psi\,, \widetilde{\Psi}\,
\mbox{and}\, \Psi$ as \br S_{NCMT_{2}}[W^{\pm},
\widetilde{W}^{\pm},{\cal W}^{\pm},\widetilde{{\cal W}}^{\pm}
]&\equiv & S[W^{\pm},\widetilde{W}^{\pm}]+S[{\cal
W}^{\pm},\widetilde{{\cal W}}^{\pm}] \nonumber\\\nonumber
 &=&\int
d^2x \{\[
 2i\widetilde{\psi}_{L}\pa_{+}\psi_{L} +
 2i\widetilde{\psi}_{R}\pa_{-}\psi_{R}-im_{\psi}\(\widetilde{\psi}_{R}\psi_{L}
-\widetilde{\psi}_{L}\psi_{R}\)-\\\nonumber
&&\l\(\widetilde{\psi}_{R}\star \psi_{R}\star
\widetilde{\psi}_{L}\star \psi_{L}+ \psi_{R}\star
\widetilde{\psi}_{R}\star \psi_{L}\star \widetilde{\psi}_{L} \)\]+\\
&&\[\psi_{R, L} \rightarrow \Psi_{R, L}, \widetilde{\psi}_{R, L}
\rightarrow \widetilde{\Psi}_{R, L}\]\} \label{ncmt2}\er which is
related to the Grisaru-Penati model (NCSG$_{2}$) defined for
$U(1)_{C}$\,fields\,
  $g\,\, \mbox{and}\,\,\bar{g}$\, of section \ref{GP}.

The Eqs. of motion comprise (\ref{ncth1})-(\ref{ncth4}) for $\psi
, \widetilde{\psi}$ and analogous Eqs. for $\Psi ,
\widetilde{\Psi}$.

In addition to the Eqs. (\ref{adeq1})-(\ref{conjugation}) for the
fields $\{{\psi},\,\widetilde{\psi}\}$,
 we must have the equations
\br \nonumber \pa_{-} (\widetilde{\Psi}_{R}\star \Psi_{R})-\pa_{+}
(\widetilde{\Psi}_{L}\star \Psi_{L})&=& \, m_{\Psi}
(\widetilde{\Psi}_{R}\star \Psi_{L}+\widetilde{\Psi}_{L}\star
\Psi_{R})-\\&& i\, \l (\widetilde{\Psi}_{R}\star \Psi_{R}\star
\widetilde{\Psi}_{L}\star \Psi_{L} - \widetilde{\Psi}_{L}\star
\Psi_{L}\star \widetilde{\Psi}_{R}\star \Psi_{R}) \label{adeq21}
 \er
and an additional equation obtained from (\ref{adeq21}) by
conveniently substituting \br \label{conjugation1}
\{\Psi_{R},\,\Psi_{L} \}\leftrightarrow
\{\widetilde{\Psi}_{R},\,\widetilde{\Psi}_{L} \}
\,\,\,\,\mbox{and}\,\,\,\, i \rightarrow -i .\er

The equations (\ref{adeq21}) and the one obtained through
 (\ref{conjugation1}) are also satisfied as the result of the relevant field
equations.

The currents conservation laws become (\ref{adeq2}) for ($\psi ,
\widetilde{\psi}$) and analogous ones for $\Psi,
\widetilde{\Psi}$, i.e. \br \label{adeq22} \pa_{-}
(\widetilde{\Psi}_{R}\star \Psi_{R})+\pa_{+}
(\widetilde{\Psi}_{L}\star \Psi_{L})&=&0 ;\,\,\,\,\,\pa_{-}
(\Psi_{R}\star \widetilde{\Psi}_{R})+\pa_{+} (\Psi_{L}\star
\widetilde{\Psi}_{L})\,=\,0 \er

However, the symmetries  associated to the currents (\ref{adeq2})
and (\ref{adeq22}) must be discussed in a correct way taking into
account the complexified $U(1)_{C}$ symmetries of its NCSG$_{2}$
sector (\ref{exp}). Thus, as in the above discussion on the
$NCSG_{1} \leftrightarrow NCMT_{1}$ case, we may attempt to
recognize the $U(1)_{C}$ group symmetries (\ref{exp}) of the
NCSG$_{2}$ model in the corresponding NCMT$_{2}$ theory. As
mentioned above in the NC case there are various ways to perform
the Noether procedure in order to obtain the currents \cite{liao}.
The $U(1)_{C}$ symmetry of a free massless fermion in commutative
space has been considered in \cite{naculich}. Following the above
discussions on the implementation of a global symmetry in the NC
case the most general  $U(1)_{C}$ transformations deserve
attention
 \br \label {u111}\psi_{R} &\rightarrow & h_{R} \star \psi_{R}\star
 g^{-1}_{R};\,\,\,\,\,\,\,\,\,\,\,\,\psi_{L} \rightarrow h_{L}\star \psi_{L}
 \star g_{L}^{-1},
\\\label{u112} \widetilde{\psi}_{R} &\rightarrow &
g_{R}\star \widetilde{\psi}_{R}\star
h_{R}^{-1};\,\,\,\,\,\,\,\,\,\,\widetilde{\psi}_{L} \rightarrow
g_{L}\star \widetilde{\psi}_{L} \star h_{L}^{-1}.\er

In (\ref{u111})-(\ref{u112}) one has \br \label{exp1}
h_{R}&=&e^{[i\l(x)-\rho(x)]};\,\,\,\,\,h_{L}=
e^{[i\l(x)+\rho(x)]};\,\,\,\,\,\,h_{R}^{\dagger}=h_{L}^{-1},\,\,\,\,
\l, \rho =\mbox{real functions}\\
g_{R}&=&e^{[i\sigma(x)-\zeta(x)]};\,\,\,\,\,g_{L}=
e^{[i\sigma(x)+\zeta(x)]};\,\,\,\,\,\,g_{R}^{\dagger}=g_{L}^{-1},\,\,\,\,
\s, \zeta =\mbox{real functions}\label{exp2} \er

In the first equation of (\ref{adeq2}) one recognizes the current
associated to the unitary $U(1)$ sector of the groups
$g_{R},\,g_{L}$ in (\ref{u111})-(\ref{u112}); i.e. the $U(1)_{C}$
symmetry provided that  $g_{R}=g_{L}$\, $(\zeta=0)$. On the other
hand, the second equation of (\ref{adeq2}) corresponds to the
other $U(1)$ symmetry related to the groups $h_{R}, \, h_{L}$ in
(\ref{u111})-(\ref{u112}) such that $h_{R}=h_{L}$\, $(\rho=0)$ .
Then the non-unitary representations  of the star-localized $U(1)$
symmetries (i.e. a representation of $U(1)_{C}/U(1)$) do not
provide conserved currents through the Noether procedure in the
NCMT$_{2}$ theory. This fact is clearly observed in the mass term
of (\ref{ncmt2}) which is not invariant under global $U(1)_{C}$
symmetries given by $h_{R} g_{R}^{-1}$ and $h_{L} g_{L}^{-1}$.
Moreover, the interaction terms are invariant under the global
symmetries but not under the localized symmetries.

Similar transformation rules can be associated to the fields
$\Psi,\,\widetilde{\Psi}$ which give rise to the conservation laws
in (\ref{adeq22}) corresponding to the other unitary subgroup of
the NC symmetry group $U(1)_{C}$ to which $\bar{g}$ is related in
the NCSG$_{2}$ model.

It is clear that the sector defined by the fields
$\psi,\,\widetilde{\psi}$ is associated to the $U(1)$ subgroup of
$U(1)_{C}$, and the $\Psi,\,\widetilde{\Psi}$ sector to a $U(1)$
subgroup of the other $U(1)_{C}$ symmetry. Then, the full
$U(1)_{C}$ symmetries of the NCSG$_{2}$ model do not provide
conserved Noether currents in the (B)NCMT$_{2}$ sector through the
process of star localizing the symmetries.

The zero-curvature condition encodes integrability even in the NC
extension of integrable models \cite{cabrera}, then we may
conclude that the (Bosonic) NCMT$_{1,2}$ theories are integrable
and infinite conserved charges may be constructed for them.

In Fig. 1 we have outlined the various relationships. Notice that
we emphasized the duality relationship NCSG$_{1}$
$\leftrightarrow$ NCMT$_{1}$ since in this case the $U(1) \mbox{x}
U(1)$ symmetry of the NCSG$_{1}$ model is implemented in the
star-localized Noether procedure to get the $U(1)$ currents of the
NCMT$_{1}$ sector.

\vskip 1.0in

\begin{picture}(11724,4854)(300,-5500)
\put(1200,-5000){\makebox(0,0)[lb]{\smash{\SetFigFont{10}{12.2}{it}Fig
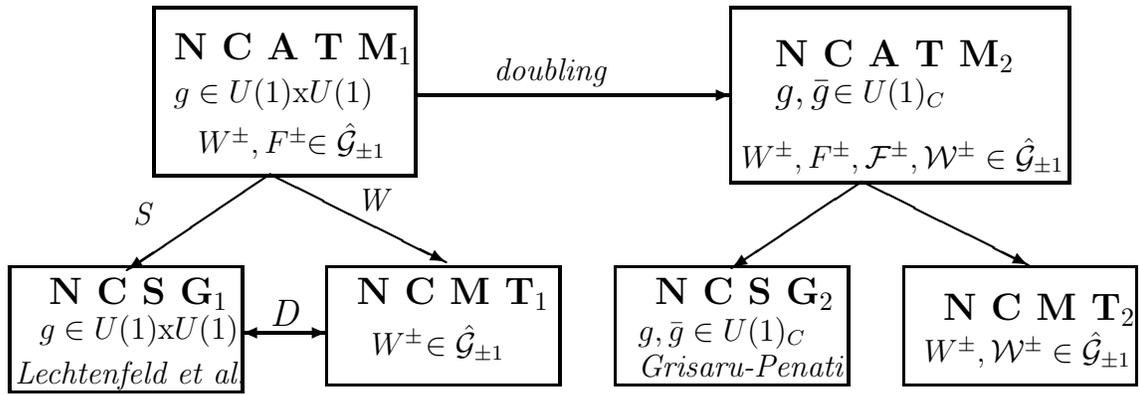
1. The models and their relationships, as well as the field
contents.}}}
\put(1900,-5250){\makebox(0,0)[lb]{\smash{\SetFigFont{10}{12.2}{it}
The duality: S=strong sector; W=weak sector; D= S-W duality}}}
\put(3301,-3661){\makebox(0,0)[lb]{\smash{\SetFigFont{14}{16.8}{it}D}}}
\put(7126,-3811){\makebox(0,0)[lb]{\smash{\SetFigFont{12}{14.4}{it}$g,
\bar{g} \in U(1)_{C}$}}}
\put(4351,-3961){\makebox(0,0)[lb]{\smash{\SetFigFont{12}{14.4}{it}$W^{\pm}$}}}
\put(4800,-3961){\makebox(0,0)[lb]{\smash{\SetFigFont{12}{14.4}{it}
$\in \hat{{\cal G}}_{\pm 1}$}}}
\put(7126,-4186){\makebox(0,0)[lb]{\smash{\SetFigFont{12}{14.4}{it}Grisaru-Penati}}}
\put(650,-4261){\makebox(0,0)[lb]{\smash{\SetFigFont{12}{14.4}{it}Lechtenfeld
et al.}}}
\put(5626,-1111){\makebox(0,0)[lb]{\smash{\SetFigFont{12}{14.4}{it}doubling}}}
\put(8200,-2011){\makebox(0,0)[lb]{\smash{\SetFigFont{12}{14.4}{}$W^{\pm},
F^{\pm}, {\cal F}^{\pm}, {\cal W}^{\pm} \in \hat{\cal{G}}_{\pm
1}$}}}
\put(3680,-1861){\makebox(0,0)[lb]{\smash{\SetFigFont{12}{14.4}{it}$\in
\hat{{\cal G}}_{\pm 1}$}}}
\put(2551,-1861){\makebox(0,0)[lb]{\smash{\SetFigFont{12}{14.4}{it}$W^{\pm},
F^{\pm}$}}}
\put(4201,-2461){\makebox(0,0)[lb]{\smash{\SetFigFont{12}{14.4}{it}W}}}
\put(7301,-3436){\makebox(0,0)[lb]{\smash{\SetFigFont{14}{16.8}{bf}N
C S G$_{2}$}}}
\put(4200,-3436){\makebox(0,0)[lb]{\smash{\SetFigFont{14}{16.8}{bf}N
C M T$_{1}$}}}
\put(8551,-961){\makebox(0,0)[lb]{\smash{\SetFigFont{14}{16.8}{bf}N
C A T M$_{2}$}}}
\put(2300,-886){\makebox(0,0)[lb]{\smash{\SetFigFont{14}{16.8}{bf}N
C A T M$_{1}$}}} \thicklines 
\put(3001,-3736){\vector(-1, 0){  0}} \put(3001,-3736){\vector( 1,
0){900}} \put(4801,-1261){\vector( 1, 0){3300}}
\put(9451,-2161){\vector( 2,-1){1740}}
\put(9451,-2161){\vector(-3,-2){1350}}
\put(3301,-2086){\vector(2,-1){1830}}
\put(3301,-2086){\vector(-3,-2){1488.461}}
\put(8101,-2161){\framebox(3500,1800){}}
\put(2101,-2086){\framebox(2700,1725){}}
\put(9901,-4261){\framebox(2400,1200){}}
\put(6901,-4261){\framebox(2400,1200){}}
\put(3901,-4261){\framebox(2400,1200){}}
\put(601,-4336){\framebox(2400,1275){}}
\put(1000,-3436){\makebox(0,0)[lb]{\smash{\SetFigFont{14}{16.8}{bf}N
C S G$_{1}$}}}
\put(900,-3800){\makebox(0,0)[lb]{\smash{\SetFigFont{12}{14.4}{}$g
\in U(1) \mbox{x} U(1)$ }}}
\put(1876,-2611){\makebox(0,0)[lb]{\smash{\SetFigFont{12}{14.4}{it}S}}}
\put(10100,-4036){\makebox(0,0)[lb]{\smash{\SetFigFont{12}{14.4}{it}$W^{\pm},
{\cal W}^{\pm} \in {\hat{\cal G}}_{\pm 1}$}}}
\put(8551,-1336){\makebox(0,0)[lb]{\smash{\SetFigFont{14}{16.8}{it}$g,
\bar{g}$}}}
\put(2300,-1336){\makebox(0,0)[lb]{\smash{\SetFigFont{12}{14.4}{}$g
 \in U(1) \mbox{x} U(1)$}}}
\put(9150,-1336){\makebox(0,0)[lb]{\smash{\SetFigFont{12}{14.4}{it}$\in
U(1)_{C}$}}}
\put(10300,-3600){\makebox(0,0)[lb]{\smash{\SetFigFont{14}{16.8}{bf}N
C M T$_{2}$}}}
\end{picture}

\section{Non-commutative solitons and strong-weak duality}
\label{ncsol}

In this section we will show that the NCATM$_{1, 2}$ models reduce
to the ordinary ATM theory in the commutative limit $\theta
\rightarrow 0$. Moreover, we will deal with the problem of the
soliton-particle and strong-weak mappings by explicitly
constructing the non-commutative solitons of the NCATM$_{1,2}$
models, respectively. In the following all field products are
understood to be $*$ products.

\subsection{NCATM$_{2}$ model and $U(1)_{C}$ parameterization}

\label{ncsg2sub}

Let us consider first the second model and the $F^{\pm}$ fields
defined by (\ref{fi2}) in the matrix representation (\ref{basis})
and $g$ given in (\ref{exp}). Then the equation (\ref{eqmg1})
becomes \br \nonumber &&\frac{-i}{m_{\psi}}\left(
\begin{array}{cc}
\pa_{-}(e^{-i \vp} \pa_{+} e^{i \vp}) &  0 \\
0 &  \pa_{-}(e^{i \vp} \pa_{+} e^{-i \vp})
\end{array} \right) =\\
&& \left(\begin{array}{cc} \psi_{L} e^{-i \vp}
\widetilde{\psi}_{R} e^{-i \vp}+ e^{i
  \vp} \psi_{R} e^{i \vp} \widetilde{\psi}_{L} &  0 \\
0 &  -\widetilde{\psi}_{L} e^{i \vp} \psi_{R}  e^{i \vp}- e^{-i
  \vp} \widetilde{\psi}_{R} e^{-i \vp} \psi_{L}
\end{array} \right)
 \label{eqnmat}
\er

Taking the trace of (\ref{eqnmat}) one can get the equation \br
\nonumber &\,&\frac{-i}{m_{\psi}}\pa_{-}(e^{-i \vp} \pa_{+} e^{i
\vp}+ e^{i \vp} \pa_{+}
e^{-i \vp})=\\
&\,&\psi_{L} e^{-i \vp}    \widetilde{\psi}_{R} e^{-i \vp}+ e^{i
  \vp} \psi_{R} e^{i \vp} \widetilde{\psi}_{L}- \widetilde{\psi}_{L} e^{i \vp} \psi_{R} e^{i \vp}- e^{-i
  \vp} \widetilde{\psi}_{R} e^{-i \vp} \psi_{L}\nonumber.
\\
\label{const} \er This equation is related to the $U(1)_{V}$
subgroup of the group $GL(2)$ related to the construction of
  NCATM$_{2}$, and reduces to a trivial equation in the limit
$\theta \rightarrow 0$.

Replacing (\ref{fi1})-(\ref{fi2}) into the equations
(\ref{eqmf11}) one can get the following system of equations \br
\label{ncspin1} \pa_{+}\psi_{L}&=& - \frac{m_{\psi}}{2} e^{i \vp}
\psi_{R} e^{i
  \vp},\,\,\,\,\,\,\pa_{+}\widetilde{\psi}_{L}\,\,=\,\, - \frac{m_{\psi}}{2} e^{-i
  \vp} \widetilde{\psi}_{R}  e^{-i \vp}\\
\pa_{-}\psi_{R}&=&  \frac{m_{\psi}}{2} e^{-i \vp} \psi_{L}  e^{-i
  \vp},\,\,\,\,\,\,\pa_{-}\widetilde{\psi}_{R}\,\,=\,\,  \frac{m_{\psi}}{2} e^{i
  \vp} \widetilde{\psi}_{L}  e^{i \vp}\label{ncspin2}
\er

Notice that the equations (\ref{eqnmat})-(\ref{ncspin2}) reduce to
the ordinary ATM equations (\ref{atm1})-(\ref{atm3}) in the limit
$\theta \rightarrow 0$.

The decoupling equations (\ref{decg1})-(\ref{decef}) provide some
relationships between the fields of the weak and strong sectors of
the NCATM$_{2}$ model. Thus, from (\ref{decg1})) one gets \br &&
\left(\begin{array}{cc} \psi_{L} e^{-i \vp} \widetilde{\psi}_{R}
e^{-i \vp}+ e^{i
  \vp} \psi_{R} e^{i \vp} \widetilde{\psi}_{L} &  0 \\
0 &  -\widetilde{\psi}_{L} e^{i \vp} \psi_{R}  e^{i \vp}- e^{-i
  \vp} \widetilde{\psi}_{R} e^{-i \vp} \psi_{L}
\end{array} \right) = \nonumber\\
&&\frac{iM^2}{m_{\psi}}\left( \begin{array}{cc}
e^{2i \vp}- e^{-2i \vp} &  0 \\
0 &  e^{-2i \vp}- e^{2i \vp}
\end{array} \right)
 \label{dec12}
\er

Taking the trace of (\ref{dec12}) one can get the equation \br
\psi_{L} e^{-i \vp} \widetilde{\psi}_{R} e^{-i \vp}+ e^{i
  \vp} \psi_{R} e^{i \vp} \widetilde{\psi}_{L}- \widetilde{\psi}_{L} e^{i \vp} \psi_{R} e^{i \vp}- e^{-i
  \vp} \widetilde{\psi}_{R} e^{-i \vp} \psi_{L}\,=\,0
\label{const11}. \er

This equation reduces to a trivial equation in the limit $\theta
\rightarrow 0$.

The remaining Eqs. (\ref{decf1})-(\ref{decf2}) give
\begin{eqnarray}
m_{\psi} e^{-i \varphi} \psi_{L}  e^{-i \varphi} &=& m_{\psi}
{\psi}_{L} - i \lambda\, \({\psi}_{R}  \widetilde{\psi}_{L}
\psi_{L}+ \psi_{L}  \widetilde{\psi}_{L} {\psi}_{R}\)\label{bil1}
\\
m_{\psi} e^{i \varphi}  \widetilde{\psi}_{L}  e^{i \varphi} &=&
m_{\psi} \widetilde{\psi}_{L} + i \lambda\, \(\widetilde{\psi}_{R}
 \psi_{L}
\widetilde{\psi}_{L}+ \widetilde{\psi}_{L}  \psi_{L}  \widetilde{\psi}_{R}\)\label{bil2}\\
 m_{\psi}
e^{i \varphi} \psi_{R}  e^{i \varphi} & = & m_{\psi} {\psi}_{R} +
i\lambda\, \( {\psi}_{L}  \widetilde{\psi}_{R}  \psi_{R}+ \psi_{R}
 \widetilde{\psi}_{R}  {\psi}_{L}\)\label{bil3}
\\
m_{\psi} e^{-i \varphi} \widetilde{\psi}_{R}  e^{-i \varphi} & = &
m_{\psi} \widetilde{\psi}_{R} - i \lambda\, \(\widetilde{\psi}_{L}
 \psi_{R}  \widetilde{\psi}_{R}+ \widetilde{\psi}_{R}
\psi_{R} \widetilde{\psi}_{L} \)\label{bil4}
\end{eqnarray}

Moreover, the Eqs. (\ref{dec12})-(\ref{bil4}) should be satisfied
by a subset of solutions of the field equations
(\ref{eqnmat})-(\ref{ncspin2}) such that the two sectors decouple.

\subsubsection{NCATM$_{2}$ one-solitons}

\label{ncatm2solitons}

It is known that the one-soliton solution of certain models solves
their NC counterparts. This feature holds in models such as the
sine-Gordon (SG) and nonlinear Schrodinger(NS) and their NC
versions NCSG$_{2}$ \cite{grisaru1, cabrera} and NCNS
\cite{Dimakis}, respectively. We will show that this behavior is
maintained in the ATM and its corresponding NCATM$_{2}$ model.
This would mean that from the point of view of one-soliton
solutions the ATM and the NCATM$_{2}$ equations of motion are the
same. Of course, the constraint (\ref{const}) must also be
verified for the decoupling sectors to have one-soliton solutions.
The study of multi-solitons is also interesting, e.g. in order to
check the validity of the NC version of the Noether and
topological currents equivalence (\ref{equiv2}).

It is also known that if $f(x_{0}, x_{1})$ and $g(x_{0}, x_{1})$
depend only on $(x_{1}-v  x_{0})$, then the product $f\star g$
coincides with the ordinary product $f.g$ \cite{cabrera, Dimakis}.
Therefore, all the $\star$ products in
(\ref{eqnmat})-(\ref{ncspin2}) for this type of functions become
the same as the ordinary ones, in this way our system of Eqs.
reduce to the usual ATM Eqs. of motion (\ref{atm1})-(\ref{atm3});
observe that e.g., $e^{i\vp}_{\star}=e^{i\vp}$, and the products
of type, $e^{i\vp}_{\star}\star \psi_{R}\star
e^{i\vp}_{\star}\star \widetilde{\psi_{L}}= e^{2i\vp}\psi_{R}
\widetilde{\psi_{L}}$.

The best example of that is the real one-soliton solution of
(\ref{atm1})-(\ref{atm3}) which was calculated in \cite{matter}
and it is given by \br
\vp &=& 2 \arctan \( \exp \( 2m_\psi \( x-x_0-vt\)/\sqrt{1-v^2}\)\) \label{solsimple1}\\
\psi &=& e^{i\theta} \sqrt{m_\psi} \, e^{m_\psi \(
x-x_0-vt\)/\sqrt{1-v^2}}\, \left(
\begin{array}{c}
\left( \frac{1-v}{1+v}\right)^{1/4}
\frac{1}{1 + i e^{2 m_\psi \( x-x_0-vt\)/\sqrt{1-v^2}}}\\
-\left( \frac{1+v}{1-v}\right)^{1/4} \frac{1}{1 - i e^{2 m_\psi \(
x-x_0-vt\)/\sqrt{1-v^2}}}
\end{array}\right)
 \label{solsimple2}
\er and the solution for $\widetilde \psi$ is the complex
conjugate of $\psi$. Thus, (\ref{solsimple1})-(\ref{solsimple2})
is a {\sl one-soliton} solution of NCATM$_{2}$. Notice that this
solution corresponds to a subset of solutions such that $g \in
U(1)$ providing the ATM model with real Lagrangian.

It is a simple observation that the solution
(\ref{solsimple1})-(\ref{solsimple2}) satisfies the relationships
(\ref{classicalboso})-(\ref{equiv2}) between the matter field
bilinears and the Toda field.

Notice that for this solution the decoupling Eq. (\ref{const11})
becomes trivial and in Eqs. (\ref{bil1})-(\ref{bil4}) one can drop
out an overall factor $\psi_{R, L}$(or  $\widetilde{\psi}_{R,L})$
in
 each equation. Thus, all of the decoupling equations  turn out to be written
 in terms of fields of type $e^{\pm 2i \vp}$ and bilinears
 $\widetilde{\psi}_{R}\psi_{L}, \widetilde{\psi}_{L}\psi_{R}$. Then, a
 relation between the NCSG$_{2}$ parameter $M$ and
the NCMT$_{2}$ coupling $\l$ can be established by using the
decoupling equations (\ref{dec12})-(\ref{bil4}) and the
relationship (\ref{classicalboso}). The matter field bilinears
expressed in terms of exponentials of the Toda field obtained in
this way for {\sl one-soliton (anti-soliton)} when conveniently
substituted into (\ref{dec12})-(\ref{bil4}) provide a relationship
between the parameters $M$ and $\l$
 \br \label{ws}\l \,=\, -\frac{m_{\psi}^2}{2 M^2}.
\er

Therefore, one can say that for this parameter relationship the
one-soliton solution (\ref{solsimple1})-(\ref{solsimple2})
satisfies the decoupling equations (\ref{dec12})-(\ref{bil4}).
This relationship is the same as the one given in \cite{nucl} in
the ordinary case. This shows that the strong-weak mapping holds
also in the NC case.

The topological charge in the NC case may be defined by \br
\label{charge} Q_{\rm topol.}= \frac{1}{\pi}
\int_{-\infty}^{+\infty} dx^{1} \frac{d \vp}{dx^{1}} \equiv
\sum_{n} \theta^n Q^{(n)}, \er where $\vp$ is the localized
``N-soliton'' solution to all orders in $\theta$. Notice that,
from (\ref{charge}), one indeed has $Q_{\rm topol.} =1$ for the
solution (\ref{solsimple1})-(\ref{solsimple2}). In particular for
the one-soliton, since this solution is exact (without expansion
in $\theta$) for the NC version, the higher order terms in
(\ref{charge}) vanish.

\subsection{NCATM$_{1}$ model and $U(1)$x$U(1)$ parameterization}

\label{ncsg1sub}

Let us consider $E_{\pm 2}$ given in (\ref{fi1}), the $F^{\pm}$
fields defined by (\ref{fi2}) in the matrix representation
(\ref{basis}) and the $U(1) \mbox{x} U(1)$ parameterization for
$g$ given in (\ref{u1u1}). Then the Eq. (\ref{eqmg1}) becomes  \br
\nonumber &&\frac{-i}{m_{\psi}}\left(
\begin{array}{cc}
\pa_{-}(e^{-i \vp_{+}} \pa_{+} e^{i \vp_{+}}) &  0 \\
0 &  \pa_{-}(e^{i \vp_{-}} \pa_{+} e^{-i \vp_{-}})
\end{array} \right) =\\
&& \left(\begin{array}{cc} \psi_{L} e^{-i \vp_{-}}
\widetilde{\psi}_{R} e^{-i \vp_{+}}+ e^{i
  \vp_{+}} \psi_{R} e^{i \vp_{-}} \widetilde{\psi}_{L} &  0 \\
0 &  -\widetilde{\psi}_{L} e^{i \vp_{+}} \psi_{R}
 e^{i \vp_{-}}- e^{-i
  \vp_{-}} \widetilde{\psi}_{R} e^{-i \vp_{+}} \psi_{L}
\end{array} \right)\nonumber\\
 \label{eqnmat0}
\er

Taking the trace of (\ref{eqnmat0}) one can get the equation \br
\nonumber &\,&\frac{-i}{m_{\psi}}\pa_{-}(e^{-i \vp_{+}} \pa_{+}
e^{i \vp_{+}}+ e^{i \vp_{-}} \pa_{+}
e^{-i \vp_{-}})=\\
&\,&\psi_{L} e^{-i \vp_{-}} \widetilde{\psi}_{R} e^{-i \vp_{+}}+
e^{i
  \vp_{+}} \psi_{R} e^{i \vp_{-}} \widetilde{\psi}_{L}- \widetilde{\psi}_{L} e^{i \vp_{+}} \psi_{R} e^{i \vp_{-}}- e^{-i
  \vp_{-}} \widetilde{\psi}_{R} e^{-i \vp_{+}} \psi_{L}\nonumber.
\\
\label{const0} \er

This equation is related to the $U(1)\mbox{x}U(1)$ sector of the
group $GL(2)$ used  to construct the NCATM$_{1}$ model, and in the
commutative limit reduces to the $U(1)_{V}$ free field equation of
motion \br \pa^{2} (\vp_{+}-\vp_{-})\, =\, 0 \label{free}\er

The equations (\ref{eqmf11}) for this parameterization provide the
system \br \label{ncspin10} \pa_{+}\psi_{L}&=& -
\frac{m_{\psi}}{2} e^{i \vp_{+}} \psi_{R} e^{i
  \vp_{-}},\,\,\,\,\,\,\pa_{+}\widetilde{\psi}_{L}= - \frac{m_{\psi}}{2} e^{-i
  \vp_{-}} \widetilde{\psi}_{R} e^{-i \vp_{+}}\\
\pa_{-}\psi_{R}&=&  \frac{m_{\psi}}{2} e^{-i \vp_{+}} \psi_{L}
e^{-i
  \vp_{-}},\,\,\,\,\,\,\pa_{-}\widetilde{\psi}_{R}=  \frac{m_{\psi}}{2} e^{i
  \vp_{-}} \widetilde{\psi}_{L} e^{i \vp_{+}}\label{ncspin20}
\er

Subtracting the diagonal elements of the matrix Eq.
(\ref{eqnmat0}) in the {\sl commutative} limit one gets \br
\pa_{+}\pa_{-} (\vp_{+}+\vp_{-}) &=& 2m_{\psi} \(\psi_{L}
\widetilde{\psi}_{R} e^{-i (\vp_{+}+\vp_{-})}+ \psi_{R}
\widetilde{\psi}_{L} e^{i (\vp_{+}+\vp_{-})}\). \label{phimm}\er

In view of the commutative limits (\ref{free}) and (\ref{phimm})
one can conclude that in this limit the system of equations for
the NCATM$_{1}$ model (\ref{eqnmat0})-(\ref{ncspin20}) reduce to
the ordinary ATM model equations of motion
(\ref{atm1})-(\ref{atm3}) if one identifies \br
\label{identification} (\vp_{+}+\vp_{-})\equiv 2\vp .\er

In this limit the free field $(\vp_{+}-\vp_{-})$ (\ref{free})
completely decouples from  the ordinary ATM model defined for the
$U(1)_{A}$ \, field \,  $\vp$ and the higher grading $\psi,
\widetilde{\psi}$ fields.

 The decoupling equations
(\ref{decg1})-(\ref{decef}) provide some relationships between the
fields of the weak-strong sectors of NCATM$_{1}$. Thus, from
(\ref{decg1})) one gets \br && \left(\begin{array}{cc} \psi_{L}
e^{-i \vp_{-}} \widetilde{\psi}_{R} e^{-i \vp_{+}}+ e^{i
  \vp_{+}} \psi_{R} e^{i \vp_{-}} \widetilde{\psi}_{L} &  0 \\
0 &  -\widetilde{\psi}_{L} e^{i \vp_{+}} \psi_{R}
 e^{i \vp_{-}}- e^{-i
  \vp_{-}} \widetilde{\psi}_{R} e^{-i \vp_{+}} \psi_{L}
\end{array} \right) = \nonumber\\
&&\frac{iM^2}{m_{\psi}}\left( \begin{array}{cc}
e^{i \vp_{+}} e^{i \vp_{-}}- e^{-i \vp_{-}} e^{-i \vp_{+}} &  0 \\
0 &  e^{-i \vp_{-}} e^{-i \vp_{+}}- e^{i \vp_{+}} e^{i \vp_{-}}
\end{array} \right)
 \label{dec120}
\er

Taking the trace of (\ref{dec120}) one can get the equation \br
\psi_{L} e^{-i \vp_{-}} \widetilde{\psi}_{R} e^{-i \vp_{+}}+ e^{i
  \vp_{+}} \psi_{R} e^{i \vp_{-}} \widetilde{\psi}_{L}- \widetilde{\psi}_{L} e^{i \vp_{+}} \psi_{R}  e^{i \vp_{-}}- e^{-i
  \vp_{-}} \widetilde{\psi}_{R} e^{-i \vp_{+}} \psi_{L}\,=\,0
\label{const110}. \er

This equation reduces to a trivial equation in the limit $\theta
\rightarrow 0$.

The remaining Eqs. (\ref{decf1})-(\ref{decf2}) give
\begin{eqnarray}\label{decncmt1}
m_{\psi} e^{-i \varphi_{+}} \psi_{L} e^{-i \varphi_{-}} &=&
m_{\psi} {\psi}_{L} - i \lambda\, \({\psi}_{R}
\widetilde{\psi}_{L} \psi_{L}+ \psi_{L} \widetilde{\psi}_{L}
{\psi}_{R}\)
\\\label{decncmt2}
m_{\psi} e^{i \varphi_{-}} \widetilde{\psi}_{L}  e^{i \varphi_{+}}
&=& m_{\psi} \widetilde{\psi}_{L} + i \lambda\,
\(\widetilde{\psi}_{R}
 \psi_{L}
\widetilde{\psi}_{L}+ \widetilde{\psi}_{L}  \psi_{L}
\widetilde{\psi}_{R}\)\\\label{decncmt3}
 m_{\psi}
e^{i \varphi_{+}} \psi_{R}  e^{i \varphi_{-}} & = & m_{\psi}
{\psi}_{R} + i\lambda\, \( {\psi}_{L} \widetilde{\psi}_{R}
\psi_{R}+ \psi_{R}
 \widetilde{\psi}_{R}  {\psi}_{L}\)
\\\label{decncmt4}
m_{\psi} e^{-i \varphi_{-}} \widetilde{\psi}_{R} e^{-i
\varphi_{+}} & = & m_{\psi} \widetilde{\psi}_{R} - i \lambda\,
\(\widetilde{\psi}_{L}
 \psi_{R}  \widetilde{\psi}_{R} + \widetilde{\psi}_{R}
\psi_{R} \widetilde{\psi}_{L} \)
\end{eqnarray}

The Eqs. (\ref{dec120})-(\ref{decncmt4}) should be satisfied by a
subset of solutions of the field equations
(\ref{eqnmat0})-(\ref{const0}) and
 (\ref{ncspin10})-(\ref{ncspin20}) such that the weak-strong sectors of NCATM$_{1}$ decouple.

\subsubsection{NCATM$_{1}$ one-solitons}

\label{ncatm2sol}

As usual, to search for one-soliton solutions we assume all the
fields depend on $(x^{1}-v x^{0})$, then from
(\ref{eqnmat0})-(\ref{const0}) one can arrive at the Eqs.
(\ref{free}) and (\ref{phimm}). Notice that for this type of
solutions one can formally remove the $\star$'s in all the
expressions. Therefore, (\ref{ncspin10})-(\ref{ncspin20}) reduce
to the corresponding
 equations of the ordinary ATM model (\ref{atm2})-(\ref{atm3}) for the identification (\ref{identification}).

 The Eq. (\ref{free}) admits
the trivial solution $\vp_{+}-\vp_{-}=0$, then in view of the
identification (\ref{identification}) one can write
$\vp_{+}=\vp_{-}=\vp$.

Moreover, in this limit the decoupling Eqs.
(\ref{dec120})-(\ref{decncmt4}) reduce to the Eqs.
(\ref{dec12})-(\ref{bil4}), respectively. Then, the {\sl
one-soliton} (\ref{solsimple1})-(\ref{solsimple2}) solves the
system (\ref{eqnmat0})-(\ref{const0}) and
(\ref{ncspin10})-(\ref{ncspin20}) for $\vp_{+}=\vp_{-}=\vp$. In
fact, for this identification basically all the discussions
regarding the one-soliton solution of section \ref{ncatm2solitons}
are valid.

Therefore one can conclude that the weak-strong relationship
(\ref{ws}) also holds in the soliton sector of the NCATM$_{1}$
model.

The N-solitons of  the NCSG$_{1}$ model have been proposed in
\cite{lechtenfeld}. In the one-soliton case it reduces to the
ordinary SG soliton. Let us emphasize that the NCATM one-soliton
is $1/2$ the NCSG$_{1}$ soliton
($\vp^{1-sol}_{SG}=2\vp^{1-sol}_{ATM} \equiv 2 \vp^{1-sol}$), this
is plausible since the ATM $\vp^{1-sol}$ soliton is interacting
with the matter fields $\psi$ and $\widetilde{\psi}$, and only in
the reduced model NCSG$_{1}$ one must envisage the true SG
(anti-soliton)one-soliton with topological numbers $\pm 1$,
provided that one considers the usual $2\pi$ normalization in
front of the integral Eq. (\ref{chargencsg1}).

\section{Conclusions and discussions}

\label{concl}

Some properties of the NC extensions of the ATM model and their
weak-strong phases described by NCMT$_{1,2}$ and NCSG$_{1,2}$
models, respectively, have been considered. The Fig. 1 summarizes
 the main relationships, as well as the field contents in each
 model. In the process of constructing the Noether currents one recognizes
the $U(1)\mbox{x} U(1)$ symmetry in both NCMT$_{1, 2}$ models (as
a subgroup of $U(1)_{C}\mbox{x} U(1)_{C}$ in the model 2). In the
$\theta \rightarrow 0$ limit we have the following: NCATM$_{1, 2}
\rightarrow$ ATM;\, NCSG$_{1,2}$ (the real sector of model $2$)
$\rightarrow$ SG(plus a free scalar in the case of model $1$)\,
and NCMT$_{1, 2}$ (one of the sectors of the model $2$)
$\rightarrow$ MT. The main result is the classical mappings
NCSG$_{1,2}$ $\leftrightarrow$ NCMT$_{1,2}$, respectively (the
real soliton sector of the models $2$). The mapping relating the
models $1$ is more promising since it is expected to hold on the
quantum level in view of the tree level results of
\cite{lechtenfeld}, regarding the nice properties in the
NCSG$_{1}$ sector, such as, factorizable and causal S-matrix. This
fact seems to be related to the fact that the whole $U(1)\mbox{x}
U(1)$ symmetries of the NCSG$_{1}$ model appear in the
localization procedure to get the Noether currents in its
NCMT$_{1}$ dual, whereas the $U(1)_{C}$ symmetry of NCSG$_{2}$ is
not recognized in this process when the NCMT$_{2}$ model is
considered since the mass terms and the interaction terms are not
invariant under the non-unitary sector $U(1)_{C}/U(1)$.

The same one-soliton solution solves both NCATM$_{1,2}$ models
(the real sector of model $2$) and the fact that this solution
depends only on ($x-v t$) allowed us to reduce the problem
basically on that of the known ATM properties, in this way
establishing the weak-strong correspondence in the NC realm.

It would be interesting to understand what actually determines the
systems to be integrable (in the sense of possessing a
factorizable and causal S-matrix) and dual to each other. A hint
in this direction, for the systems NCSG$_{1}$ $\leftrightarrow$
NCMT$_{1}$, seems to be that the models originate directly through
a reduction processes starting from the NC WZNW type action for
the Toda field coupled to the higher grading matter fields
(NCATM$_{1}$), such that the $U(1)\mbox{x} U(1)$ symmetry is
relevant, both in the construction of the NCSG$_{1}$ model and in
the star-localized Noether procedure to construct the $U_{1,2}(1)$
currents in the NCMT$_{1}$ sector. Recall that the commutative
integrable ATM models are themselves derivable from the (two-loop)
WZNW model \cite{nucl, matter} when a corresponding Hamiltonian
reduction is performed \cite{feher}.

Various aspects of the models studied above deserve attention in
future research, e.g., the NC multi-solitons of the NCATM$_{1,2}$
models, the bosonization of the NCMT$_{1,2}$ models and its
multi-fermion extensions \cite{jmp}, the NC zero-curvature
formulation of the MT model defined for Grassmannian fields. In
particular, in the bosonization process of NCMT$_{1, 2}$ models
and their multi-fermion extensions, initiated in  \cite{nunez} by
directly starring the usual Thirring interaction, we believe that
a careful understanding of the star-localized NC Noether
symmetries, as well as the classical soliton spectrum would be
desirable.

{\sl Acknowledgments}

HB thanks IMPA, CBPF, ICET and Prof. M. C. Ara\'ujo at the
Mathematics Department-UFMT for hospitality. The authors thank I.
Cabrera-Carnero for discussions and M. Hamanaka for correspondence
and valuable comments. HLC has been supported by CLAF and HB by
CNPq-FAPEMAT.

                \appendix

 \section{Affine Lie algebra $\hat{sl}(2)$}
 \label{appa}

The Lie algebra $sl(2)$ is formed by all $2 \times 2$ complex
matrices with zero trace.
 Let us assume the basis
\begin{equation}
H = \left(\begin{array}{cr}
1 &  0 \\
0 & -1
\end{array} \right),\,\,\,\,
E_{+} = \left(\begin{array}{cc}
0 & 1 \\
0 & 0
\end{array} \right),\,\,\,\,
E_{-} = (E_{+})^T. \label{basis}
\end{equation}
We use the invariant bilinear form on $sl(2)$ defined by \br
\label{trazo} <x\,,\,y>\equiv <x \, y> = \mbox{Tr}(xy), \qquad x,
y \in sl(2). \er

The affine Kac-Moody algebra $\widehat{sl}(2)$ is constructed as
follows. Consider the loop algebra
\begin{equation}
{\mathcal L}(sl(2)) = {\cal C}[\zeta, \zeta^{-1}] \otimes sl(2),
\end{equation}
where ${\cal C}[\zeta, \zeta^{-1}]$ is the algebra of Laurent
polynomials in $\zeta$. An element of ${\mathcal L}(sl(2))$ is a
finite linear combination of the elements of the form $\zeta^m
\otimes x$, where $m \in
 \IZ$ and $x \in sl(2)$. The structure of a Lie
algebra in ${\mathcal L}(sl(2))$ is introduced by the relation
\begin{equation}
\sbr{\zeta^m \otimes x}{\zeta^n \otimes y} = \zeta^{m+n} \otimes
\sbr{x}{y}.
\end{equation}
The elements of the form  $1 \otimes x$, $x \in  sl(2)$ are
identified with the Lie algebra $sl(2)$. Thus the algebra $sl(2)$
is  a subalgebra of ${\mathcal L}(sl(2))$. This allows us to write
$\zeta^m \otimes x $ in the form  $\zeta^{m} x$.

Let us denote by  $\tilde{\mathcal L}(sl(2))$ the algebra extended
by the one dimensional center operator $C$. The commutation
relations in the Lie algebra $\tilde{\mathcal L}(sl(2)) =
{\mathcal L}(sl(2)) \oplus {\cal C} \, C$ are given by \br
 \lb H^m \, , \, H^n \rb &=& 2 \, m \, C \, \d_{m+n,0},  \label{sl2a}\\
                \lb H^m \, , \, E^n_{\pm} \rb &=& \pm 2 \, E^{m+n}_{\pm},
                \label{sl2b}\\
   \lb E^m_{+} \, , \, E^n_{-} \rb &=& H^{m+n} + m \, C \, \d_{m+n,0},
                \label{sl2c}
  \er
where the elements $H^m = \zeta^m H,\, E_{\pm}^m = \zeta^m
E_{\pm}$ have been defined.

Next, denote by $\widehat{sl}(2)$ the Lie algebra which is
obtained by adjoining to $\tilde{\mathcal L}(sl(2))$ a derivation
operator $D = \zeta (\mathrm d/\mathrm d \zeta)$. The commutation
relations for the Lie algebra $\widehat{sl}(2)$ are defined by
relations (\ref{sl2a})-(\ref{sl2c}) and by the equalities
\begin{equation}
\sbr{D}{\zeta^m x} = m \, \zeta^m x, \qquad \sbr{D}{C} = 0.
\end{equation}

                The generator  $Q\equiv    \h H^0 + 2 D$ is the principal gradation operator \cite{kac}. Then its eigensubspaces are
\br
\hat{{\cal G}}_{0} &=& \{ H^0, C, Q\} ;\nonumber\\
\hat{{\cal G}}_{2n+1} &=& \{ E_{+}^n , E_{-}^{n+1}\} \, \qquad n\in \IZ ;\nonumber\\
\hat{{\cal G}}_{2n} &=& \{ H^n\} ,\, \qquad n\in \{ \IZ - 0\} .
\label{eigensl2} \er

                \section{The commutative $sl(2)$  ATM model}

                \label{appb}

                The $sl(2)$  ATM theory has been studied from different points of view
                \cite{nucl, nucl1, tension, annals}. Here we
                present some results expressed in matrix form which are relevant to our discussions.
                The theory contains the usual sine-Gordon (SG)
                and the massive Thirring (MT) models describing the soliton/particle correspondence of its spectrum \cite{nucl, nucl1}.

                The equations of motion of the $\hat{sl}(2)$ conformal affine
                Toda model    coupled to matter (CATM) are given by
\br \label{aeqmg1} \pa_{-}(g^{-1} \pa_{+} g) + \pa_{-}\pa_{+} \nu
C & =& e^{\eta} \[F^{-}\,,\, g F^{+} g^{-1}\]\\
\label{aeqmf11}
\partial_{+} F^{-}& =& - e^{\eta} [E_{-2}\,,\,  g  F^{+} g^{-1} ],
\,\,\,\,\,\,\,\,\,\,\,
\partial_{-} F^{+} = e^{\eta} [E_{2}\,,\,  g^{-1} F^{-} g ],\\
\partial_{-}\partial_{+}\eta Q & =&0,
\er where

\br \label{fi1} E_{\pm 2}&=& \frac{m_{\psi}}{4} H^{\pm
1},\,\,\,\,\,\,\,\, g=e^{i\vp H^{0}},\\\label{fi2} F^{+}&=&
\sqrt{i m_{\psi}} \( \psi_{R}E_{+}^{0}+ \widetilde{\psi}_{R}
E_{-}^{1} \),\,\,\,\, F^{-}= \sqrt{im_{\psi}} \(\psi_{L}
E_{+}^{-1}-\widetilde{\psi}_{L} E_{-}^{0}\)\er

We have denoted by $H^n$, $E_{\pm}^n$, and $C$ the Chevalley
  basis generators of the $\hat{sl}(2)$ affine Kac-Moody algebra and $Q$ the
  principal gradation generator.

          Taking into account the parameterizations (\ref{fi1})-(\ref{fi2}) and
             by setting $\eta =0$ one gets
                the off-critical $sl(2)$ ATM model equations of motion from (\ref{aeqmg1})-(\ref{aeqmf11})\footnote{Consider
                $\gamma_0 = -i \(
                \begin{array}{rr} 0&-1\\ 1&0
                \end{array}\)$ ,
                $\gamma_1 = -i \(
                \begin{array}{rr}  0&1\\  1&0
                \end{array}\)$,
                $\gamma_5 = \gamma_0\gamma_1 = \(
                \begin{array}{rr} 1&0\\ 0&-1
                \end{array}\)$}

\br\label{atm1} \pa_{+}\pa_{-} \vp &=&m_{\psi} \(\psi_{L}
\widetilde{\psi}_{R} e^{-2i \vp}+
\psi_{R} \widetilde{\psi}_{L} e^{2i \vp}\)\\
\label{atm2} \pa_{+}\psi_{L}&=& - \frac{m_{\psi}}{2} \psi_{R}
e^{2i
  \vp},\,\,\,\,\,\,\pa_{+}\widetilde{\psi}_{L}= - \frac{m_{\psi}}{2} \widetilde{\psi}_{R} e^{-2i \vp}\\
\pa_{-}\psi_{R}&=&  \frac{m_{\psi}}{2} \psi_{L} e^{-2i
  \vp},\,\,\,\,\,\,\pa_{-}\widetilde{\psi}_{R}=  \frac{m_{\psi}}{2} \widetilde{\psi}_{L} e^{2i \vp}
  \label{atm3}
\er

Associated to these equations of motion one can write the ATM
Lagrangian\footnote{Define the Dirac fields as \br
 \psi = \( \begin{array}{c} \psi_R\\ \psi_L  \end{array}\) \, ; \qquad \widetilde \psi = \( \begin{array}{c}
                \widetilde \psi_R\\
                \widetilde \psi_L
                \end{array}\)\, ; \,\,\,\, \bar{\psi}=\widetilde{\psi}^{T}
                \g^{0}.\er
We are using \br x_{\pm}=t\pm x,\, \mbox{then},\,
\pa_{\pm}=\frac{1}{2}(\pa_{t}\pm \pa_{x}),\, \mbox{and}\,\,
\pa^{2}=\pa^{2}_{t}-\pa_{x}^{2}=4\pa_{-}\pa_{+}.\er} \br
  {\cal L}_{ATM} =-\frac{1}{4} \pa_{\mu} \vp \, \pa^{\mu} \vp
                + i  {\bar{\psi}} \gamma^{\mu} \pa_{\mu} \psi
                - m_{\psi}\,  {\bar{\psi}} \,
                e^{2i\vp\,\gamma_5}\, \psi.   \label{atm}
\er

Notice that (\ref{atm}) is a real Lagrangian if $\vp$ is real and
$\widetilde{\psi}$ is the complex conjugate of $\psi$.

 The strong/weak dual phases of the model (\ref{atm}) has been uncovered   by means of the symplectic and master Lagrangian approaches
                \cite{annals, jhep}. The strong phase is described by the SG model: $ {\cal L}_{SG}=\frac{1}{2}(\partial_{\mu}\vp)^{2}  + \frac{m^2_{\psi}}{\l} \; \mbox{cos} 2 \vp$.
                The usual massive Thirring model ${\cal L}_{MT}=  i\overline{\psi}\gamma^{\mu}\pa_{\mu}\psi-m_{\psi}\overline{\psi}\psi+
                \frac{1}{2} \l j_{\mu}j^{\mu}$ describes the weak coupling phase.  Notice the weak-strong exchange $\l \rightarrow \frac{1}{\l}$ of the
                coupling constant.

The one-(anti)soliton solution of the system
(\ref{atm1})-(\ref{atm3}) satisfies the remarkable SG and MT
classical
                correspondence \cite{orfanidis} in which,  apart from the Noether and
                topological currents equivalence, MT matter field  bilinears
                are related to the exponentials of the SG field \cite{nucl}.

                The soliton type solutions satisfy the relationships \cite{nucl}         \br
  \label{classicalboso}
  \psi _{R}\tilde{\psi}_{L}=\frac{im_{\psi}}{2\l}(e^{-2 i\varphi }-1),\,\,\,\,&&\psi _{L}  \tilde{\psi}_{R}=-\frac{im_{\psi}}{2\l}(e^{2i\varphi }-1),
 \er
         and the currents equivalence
  \br
  \label{equiv2}
 \bar{\psi}\gamma^{\mu }\psi&=&\h
   \epsilon^{\mu\nu}\pa_{\nu}\vp.
   \er

                The relationships (\ref{classicalboso}) and (\ref{equiv2}) have been
                verified for $N=1$ and $N=1, 2$ solitons, respectively \cite{nucl, nucl1}.

In order to write the matrix form of the Lagrangian let us
consider the auxiliary fields $ W^{\mp}$ defined by \br
 \partial_{+} W^{+} =  -g  F^{+}  g^{-1},\,\,\,\,
  \partial_{-} W^{-} = - g^{-1}  F^{-} g, \label{w2},
\er

Next, setting $\eta=0$ in (\ref{aeqmg1})-(\ref{aeqmf11}) one can
write the matrix form of the off-critical ATM action \cite{jhep}
\br \nonumber S[g, W^{\pm}, F^{\pm}] &=& I_{WZW}[g] + \int
\{\frac{1}{2}<
\partial_{-} W^{-} [E_{2}\,,\, W^{-} ]> -\frac{1}{2}
< [E_{-2}\,,\,W^{+}] \partial_{+}
                W^{+} >\\
&& + < F^{-} \partial_{+} W^{+} > + <\partial_{-} W^{-}
                 F^{+}>+ <F^{-} g F^{+} g^{-1}  > \},
\label{atmmat} \er where $I_{WZW}[g]$ is the Wess-Zumino-Witten
model. The notation $<,>$ extends the trace operation
(\ref{trazo}) to the affine case. Notice that since $g$ is Abelian
the WZW term in $I_{WZW}$ vanishes.

\end{document}